\documentstyle[aas2pp4]{article}

\slugcomment{Accepted for publication in ApJ}

\lefthead{Simard et al.}
\righthead{Magnitude$-$Size Relation of Galaxies}

\def\deg{\mbox{$^\circ$}}

\def\littleprime{\ifmmode{\scriptscriptstyle \prime } 
\else{\hbox{$\scriptscriptstyle \prime$ }}\fi}

\def\arcsec{\raise .9ex \hbox{\littleprime\hskip-3pt\littleprime}}
\def\arcmin{\raise .9ex \hbox{\littleprime}}
\def\arcsecpoint{\hbox to 1pt{}\rlap{\arcsec}.\hbox to 2pt{}}
\def\arcminpoint{\hbox to 1pt{}\rlap{\arcmin}.\hbox to 2pt{}}

\def\spose#1{\hbox to 0pt{#1\hss}}
\def\lta{\mathrel{\spose{\lower 3pt\hbox{$\mathchar"218$}}
     \raise 2.0pt\hbox{$\mathchar"13C$}}}
\def\gta{\mathrel{\spose{\lower 3pt\hbox{$\mathchar"218$}}
     \raise 2.0pt\hbox{$\mathchar"13E$}}}

\begin{document}

\title{The Magnitude-Size Relation of Galaxies out to z $\sim$ 
1\altaffilmark{1,2}}

\author{Luc Simard\altaffilmark{3}, David C. Koo, S. M. Faber, 
Vicki L. Sarajedini, Nicole P. Vogt\altaffilmark{3,4}, Andrew C. 
Phillips, Karl Gebhardt, Garth D. Illingworth, K. L. Wu} 
\affil{UCO/Lick Observatory, University of California, Santa Cruz, CA 
95064, USA}

\altaffiltext{1}{Based on data obtained at 
the W.M. Keck Observatory, which is operated as a scientific 
partnership among the California Institute of Technology, the 
University of California and the National Aeronautics and Space 
Administration.  The Observatory was made possible by the generous 
financial support of the W.M. Keck Foundation.}

\altaffiltext{2}{Based on observations made with the NASA/ESA {\it 
Hubble Space Telescope} which is operated by AURA, Inc., under 
contract with NASA.}

\altaffiltext{3}{Guest User, Canadian Astronomy Data Center, which is 
operated by the National Research Council of Canada, Herzberg 
Institute of Astrophysics, Dominion Astrophysical Observatory} 

\altaffiltext{4}{Current Address: Institute of Astronomy, University 
of Cambridge, Cambridge CB3 0HE, England, UK}

\begin{abstract}
  As part of the Deep Extragalactic Evolutionary Probe (DEEP) survey, 
  a sample of 190 field galaxies ($I_{814} \leq 23.5$) in the ``Groth 
  Survey Strip'' has been used to analyze the magnitude-size relation 
  over the range $0.1 < z < 1.1$.  The survey is statistically 
  complete to this magnitude limit.  All galaxies have photometric 
  structural parameters, including bulge fractions ($B/T$), from 
  Hubble Space Telescope images, and spectroscopic redshifts from the 
  Keck Telescope.  The analysis includes a determination of the survey 
  selection function in the magnitude-size plane as a function of 
  redshift, which mainly drops faint galaxies at large distances.  Our 
  results suggest that selection effects play a very important role.
  
  A first analysis treats disk-dominated galaxies with $B/T < 0.5$.  
  If selection effects are ignored, the mean disk surface brightness 
  (averaged over all galaxies) increases by $\sim$ 1.3 mag from $z$ = 
  0.1 to 0.9.  However, most of this change is plausibly due to 
  comparing low luminosity galaxies in nearby redshift bins to 
  high luminosity galaxies in distant bins. If this effect is allowed 
  for, no discernible evolution remains in the disk surface brightness 
  of bright ($M_{B} < -19$) disk-dominated galaxies. A second analysis 
  treats all galaxies by substituting half-light radius for disk scale 
  length, with similar conclusions. Indeed, at all redshifts, the bulk 
  of galaxies is consistent with the magnitude-size envelope of local 
  galaxies, i.e., with little or no evolution in surface 
  brightness. In the two highest redshift bins ($z > 0.7$), a handful 
  of luminous, high surface brightness galaxies appears that occupies 
  a region of the magnitude-size plane rarely populated by local 
  galaxies. Their wide range of colors and bulge fractions points to 
  a variety of possible origins.
  
\end{abstract}


\section{Introduction} \label{intro}
The field of galaxy evolution is rapidly changing under the wealth of 
new data coming from powerful instruments such as the Hubble Space 
Telescope ({\it HST}) and 10-m class telescopes.  Given the complexity of 
the problem, any attempt at understanding galaxy evolution must be 
deeply rooted in a quantitative and systematic study of fundamental 
galaxy structural parameters such as luminosity, size, mass {\it and} 
their associated scaling relations.  High-redshift galaxies were first 
studied through their global luminosity function.  The landmark 
Canada-France Redshift Survey (CFRS, \cite{lilly95}) showed that the 
luminosity function of blue galaxies evolves significantly over the 
redshift range 0.0 $< z <$ 1.0 while that of red galaxies shows little 
or no evolution over the same redshift range.  However, luminosity 
functions cannot paint a complete picture of the evolution of 
galaxies, as luminosity is only one of many galaxy structural 
parameters that may be evolving.

Since individual galaxies cannot be observed at different epochs, 
galaxy evolution can be operationally defined as the statistical 
change in galaxy properties with redshift.  Correlations between 
observables such as total absolute magnitude $M$, half-light radius 
$R_{hl}$, disk scale length $R_{d}$, and rotation speed $V_{c}$ are 
particularly powerful diagnostics, as they involve key global 
properties of galaxies, namely luminosity, size, and mass.  
Furthermore, shifts in the zeropoints of such correlations are fairly 
insensitive to the local densities of objects, and co-moving volume 
densities are therefore not required.

So far, two galaxy scaling relations have been explored at moderate 
redshifts.  The $M-V_{c}$ relation, also more commonly known as the 
Tully-Fisher (TF) relation, has been studied to $z \sim 1$ by Vogt 
et al. (1996,1997).  Based on only 16 galaxies, they found that: 
(1) massive, mature disks were present at $z \sim 1$, and (2) the 
change in the TF B-band zeropoint of these disks was at most 0.4 mag.  
Using ground-based CCD images, Schade et al.  (1996a,b) presented 
$R_{d}-M_{disk}$ relations over the redshift range $0.1 < z < 0.6$ for 
351 late-type field and cluster galaxies and $R_{e}-M_{bulge}$ 
relations for 166 early-type field and cluster galaxies over the same 
redshift range.  Both samples were taken from the ground-based Canadian 
Network for Observational Cosmology (CNOC) sample.  The late-type 
galaxy magnitude-size relations showed an apparent increase in disk 
surface brightness of $\sim$ 1 mag at a given $R_d$ by $z = 0.6$, both 
in the outer regions of clusters and in the field.  Similarly, the 
effective surface brightness of early-type galaxies seemed to increase 
by 0.7 mag at $z = 0.6$.  Using {\it HST} data, Roche et al.  (1998) studied 
the surface brightness and size of 270 galaxies out to $z = 4$ taken 
from various redshift surveys.  They found that the mean rest-frame 
blue-band surface brightness at a fixed radius seemed to increase 
smoothly by 0.95 mag between $z$ = 0.2 and $z$ = 0.9.  They also 
concluded that a Pure Luminosity Evolution (PLE) model was not 
sufficient to explain the observed evolution in surface brightness and 
proposed a Size-Luminosity Evolution (SLE) model based on the 
evolution of different stellar populations as a function of 
galactocentric distance.  Lilly et al.  (1998) concluded from {\it HST} 
structural parameters of 341 CFRS and LDSS galaxies that the increase 
in the disk surface brightness with redshift for galaxies with disk 
scale lengths greater than 4 $h_{50}^{-1}$ kpc was 0.8 $\pm$ 0.3 mag.  
This increase was not sufficient to account for the evolution of the 
overall luminosity function of galaxies seen in the CFRS survey over 
the interval $0 < z < 1$.  They suggested that smaller galaxies were 
responsible for most of the observed evolution, but they could not 
conclusively verify that hypothesis due to their magnitude and 
radius limits.

The question of whether the apparent evolution seen in galaxies at 
high redshifts is operating in both luminosity and/or size has 
received considerable attention in a recent flurry of theoretical 
activity aimed at explaining the formation and evolution of disk 
galaxies (\cite{bouwens97}; \cite{dalc97}; \cite{mo98}; \cite{mao98}; 
\cite{contardo98}; \cite{firmani98}).  Certain semi-analytical 
no-infall models (e.g., \cite{mo98}) predict that disk scale length 
should vary as $(1+z)^{-1}$ in an $\Omega = 1$ universe.  Smaller disk 
scale lengths would be expected in the past due to the increase in the 
critical density $\rho_{crit}$ of the Universe with look-back time.  
Dark matter halos are defined with a fixed overdensity 
($\rho_{mean}^{halo} \sim 200\rho_{crit})$ in numerical simulations, 
and the radius of those halos must therefore decrease with increasing 
redshift to meet the overdensity criterion at high redshifts.  
Assuming that the mass and angular momentum of the gas settling into 
disks are fixed fractions of the mass and angular momentum of dark 
matter halos, smaller halo radii will translate into smaller disk 
scale lengths.  Infall models (e.g., \cite{chiappini97}), on which the 
SLE model of Roche et al.  (1998) is based, predict a milder disk size 
evolution.  Disk sizes at $z \sim 1$ in these models are typically 
only 20$\%$ smaller than at $z = 0$.  Disk size is therefore important 
by itself to distinguish between these different scenarios even 
though, as pointed out by Mao, Mo, \& White (1998), it is likely that 
only kinematic information can decisively separate evolution in the 
mass and size of galaxies from evolution in their stellar 
mass-to-light ratio.

One of the goals of the Deep Extragalactic Evolutionary Probe (DEEP; 
\cite{koo98}) survey is to compare the full 3D distribution of galaxy 
size, luminosity and rotation velocity at high redshift to local 
relations (e.g., \cite{burst97}; hereafter BBFN) using a deep, 
homogenous sample of galaxies observed with the Hubble Space Telescope 
($HST$) and the 10-m W.M. Keck Telescope.  As a first step towards 
this goal, this paper focuses on the galaxy magnitude-size relation 
and seeks to extend previous work to fainter disk luminosities and 
higher redshifts.  A major new focus is the role of selection effects.  
As this paper will show, selection effects have had a major impact on 
the conclusions drawn from previous studies of the magnitude-size 
relation.  Further photometric data, internal velocities and full 3D 
structural distributions on larger samples of DEEP galaxies will 
follow in future papers.

Section~\ref{obsanal} describes the sample and observational 
measurements of galaxy sizes and magnitudes.  Selection effects are 
quantified in Section~\ref{seleff}.  The magnitude-size relation as a 
function of redshift is presented in Section~\ref{sizemag}, and its 
implications for galaxy evolution are discussed in 
Section~\ref{discuss}.  Values of $H_{0}$ = 70 km s$^{-1}$ Mpc$^{-1}$ 
and $q_{0} = 0.1$ are used throughout this paper so that $L^*$ 
galaxies have $I \sim 22.9$ at $z \sim 1$ with 1$\arcsec$ 
corresponding to 7.3 kpc.  It should be noted that surface brightness 
is independent of $q_{0}$. All magnitudes are on the Vega-based 
system ($I_{814,Vega} = I_{814,AB}-0.44$).

\section{Observations and Analysis} \label{obsanal} 

\subsection{HST Structural Measurements} \label{hstim}

The {\it HST} data are from two surveys, which we dub the ``Groth Survey 
Strip'' (GSS), taken under {\it HST} programs GTO 5090 and GTO 5109.  The GSS 
consists of 28 overlapping subfields taken with the {\it HST} Wide Field and 
Planetary Camera (WFPC2) and forms a ``chevron strip'' oriented NE to 
SW at roughly 1417+52 at Galactic latitude $b \sim 60\deg$.  Each of 
27 subfields has exposures of 2800 s in the broad $V$ filter 
($F606W$) and 4400 s in the broad $I$ filter ($F814W$).  The 28th 
field is the Deep Field (J2000 1417.5+52.5), with total exposures of 
24,400 s in $V$ and 25,200 s in $I$.  The images were recalibrated 
``on-the-fly'' through the Canadian Astronomy Data Center (CADC) 
standard pipeline and coadded with the IRAF\footnote{IRAF is 
distributed by the National Optical Astronomy Observatories, which are 
operated by the Association of Universities for Research in Astronomy, 
Inc., under cooperative agreement with the National Science 
Foundation.}/STSDAS task CRREJ to remove cosmic rays.

Object catalogs were produced with SExtractor version 1.0a 
(\cite{bertin96}) using a detection threshold of 1.5$\sigma$ and a 
minimum detection area of 10 contiguous pixels.  SExtractor also 
produced a ``segmentation'' or ``mask'' image which was used to 
deblend galaxies from their nearby companions.  Pixels belonging to 
the same object are all assigned the same flag in this segmentation 
image, so the SExtractor segmentation image was ideal for isolating 
objects in the surface brightness profile fits.

The surface brightness profiles of galaxies in the DEEP/GSS sample 
were fitted with a PSF-convolved 2D bulge+disk model (GIM2D: \cite{simard98}, 
\cite{marleau98}).  The model had a {\it maximum} of twelve 
parameters: the flux $F_{total}$ in data units (DU) integrated out to 
$r = \infty$, the bulge fraction $B/T$ ($\equiv 
F_{bulge}/F_{total}$), the bulge effective radius $r_e$, the 
bulge ellipticity $e$ ($e \equiv 1-b/a$, $b \equiv$ semi-minor axis, 
$a \equiv$ semi-major axis), the bulge position angle of the major 
axis $\phi_{b}$, the exponential disk scale length $r_d$, the disk 
inclination $i$ (face-on $\equiv$ 0), the disk position angle 
$\phi_d$, the subpixel $dx$ and $dy$ offsets of the galaxy center, the 
background level $b$, and the S\'ersic index $n$.

The first component (the ``bulge'') of the 2D surface brightness model  
was a S\'ersic profile of the form:

 \begin{equation}
 	\Sigma(r) = \Sigma_{e} {\rm exp} \{-k[(r/r_{e})^{1/n} - 1]\}
 	\label{sersic}
 \end{equation} 

\noindent where $\Sigma(r)$ is the surface brightness at $r$ along the 
semi-major axis in linear flux units per unit area, and $\Sigma_{e}$ 
is the effective surface brightness.  The parameter $k$ was set equal 
to 1.9992$n$ $-$ 0.3271 so that $r_{e}$ remained the projected radius 
enclosing half of the light in this component.  The classical de 
Vaucouleurs profile has the special value $n$ = 4, and this value was 
chosen for the current analysis.  This choice was motivated by studies 
of bulge profiles in local galaxies.  Locally, there is evidence that 
the bulges of late-type spiral galaxies may be better fitted by an $n$ 
= 1 profile, whereas bright ellipticals and the bulges of early-type 
spiral galaxies follow an $n$ = 4 profile (\cite{dejong94}, 
\cite{courteau96}, \cite{andredakis98}).  Local late-type galaxies 
with $n$ = 1 bulges have $B/T \leq 0.1$ (\cite{dejong94}).  Since such 
bulges contain only 10\% of the total galaxy light, low 
signal-to-noise measurements of late-type high-redshift galaxies make 
it very difficult to determine the S\'ersic index.  On the other hand, 
$n$ is more important for bulge-dominated galaxies, and $n$ = 4 is the 
expected value based on early-type local galaxies.  The total flux in 
the S\'ersic bulge component is calculated by integrating 
Equation~\ref{sersic} from $r$ = 0 to infinity to obtain:

\begin{equation}
     F_{bulge} = 2 \pi n e^{k} k^{-2n} r_e^2 \Gamma(2n) \Sigma_{e}
     \label{sersicflux}
\end{equation}

\noindent where $\Gamma$ is the incomplete gamma function. For 
$n=4$, $F_{bulge} = 7.214 \pi r_{e}^{2} \Sigma_{e}$. 

The second component (the ``disk'') was a simple exponential profile of 
the form:

\begin{equation}
	\Sigma(r) = \Sigma_{0} {\rm exp} (-r/r_d), 
	\label{disk}
\end{equation}

\noindent where $\Sigma_{0}$ is the central surface brightness. The total 
flux in the disk is given by:

\begin{equation}
	F_{disk} = 2 \pi r_d^2 \Sigma_{0}.
	\label{diskflux}
\end{equation}

\noindent A PSF-deconvolved half-light radius $r_{hl}$ was also 
computed for each galaxy by integrating the sum of 
Equations~\ref{sersic} and~\ref{disk} out to infinity with the best 
fitting structural parameters.  Even though the conventional 
``bulge/disk'' nomenclature has been adopted in this paper, it should 
be kept in mind that this nomenclature does not say anything about the 
internal kinematics of the components.  The presence of a ``disk'' 
component does not necessarily imply the presence of an actual disk 
since many dynamically hot systems also have simple exponential 
profiles of the form given by Equation~\ref{disk} 
(\cite{lin83}, \cite{kormendy85}).  Likewise a ``bulge'' may represent 
a brightened center due to a starburst rather than a genuine 
dynamically hot spheroid.

WFPC2 detector undersampling was taken into account by generating the 
surface brightness model on an oversampled grid, convolving it with 
the appropriate point spread function (PSF), shifting its center 
according to $dx$ and $dy$, and rebinning the result to the detector 
resolution for direct comparison with the observed galaxy image.  The 
PSF was generated by the Space Telescope package {\sl TINY TIM} 
(\cite{krist93}) and subsampled to reproduce the pixel resolution of 
the Wide-Field camera.  The best fitting parameter values were found 
with the Metropolis Algorithm (\cite{metro53}, \cite{saha94}), which 
Monte-Carlo samples parameter space to maximize the likelihood 
function.  Best parameter values and their confidence intervals were 
determined simultaneously by the algorithm as it refined its best 
solution.  Figure~\ref{gim2d-orig-res} shows a 55\arcsec $\times$ 
45\arcsec section of a typical WFPC2/F814W GSS image before and after 
GIM2D processing.  Figure~\ref{gim2d-fits-examples} shows examples of 
GIM2D surface brightness fits for a set of five galaxies covering a 
wide range of bulge fractions and redshifts. 

Following the simulation procedure described in Section 3.4 of Marleau 
\& Simard (1998), 400 simulated GSS galaxies were created with a wide 
range of structural parameters and analyzed in exactly the same way as 
real galaxies to test the reliability of the GSS parameter values 
measured with GIM2D. For simulated disk-dominated galaxies ($B/T \leq 
0.5$) brighter than $I_{814}$ = 23.5, the mean difference between the 
measured and input disk magnitudes $<\Delta I_{814} (disk)>$ was $-$0.02 
with $\sigma(\Delta I_{814} (disk))$ = 0.20.  For the same galaxies, 
the mean difference between the measured and input log disk scale 
lengths $<\Delta log r_{d}>$ was 0.02, and $\sigma(\Delta log r_{d})$ 
was 0.05. Errors in bulge+disk fits are discussed in greater details in 
Marleau \& Simard (1998).

\subsection {Keck/LRIS Spectroscopy} \label{keckspec}

GSS galaxies are currently being systematically surveyed by the DEEP 
team using the multi-object Low Resolution Imaging Spectrograph (LRIS, 
\cite{oke95}) on the W.M. Keck II Telescope.  DEEP/GSS galaxies were 
selected according to the magnitude criterion $(V+I)/2 < 24$.  Spectra 
of a first group of 231 galaxies were acquired between May 1995 and 
May 1997, and these are the galaxies used in this paper.  Two gratings 
were used to cover a total spectral range of about 4500--9100 
\AA\thinspace\thinspace depending on the exact position of the target 
on the mask: a 900 lines mm$^{-1}$ grating (central wavelength 
$\simeq$ 5800 \AA, dispersion = 0.85 \AA/pixel and resolution $\sim$ 
3--4 \AA\thinspace FWHM), and a 600 lines mm$^{-1}$ grating (central 
wavelength $\simeq$ 7700 \AA, dispersion = 1.26 \AA/pixel and 
resolution $\sim$ 4--5 \AA\thinspace FWHM).  Typical total exposure 
time per target and per grating was 2700 seconds.  Rectified, 
wavelength-calibrated, sky-subtracted 2D spectra were produced with a 
custom LRIS reduction package.  One-dimensional spectra were optimally 
extracted using the IRAF/APEXTRACT package.  Details of the spectral 
data reduction will be given elsewhere.  The final DEEP/GSS sample has 
a total of 217 objects with both Keck redshifts and $HST$ structural 
parameters.  The sample was cut down further to 190 galaxies by 
selecting only galaxies brighter than $I_{814}$ = 23.5.  The DEEP/GSS 
redshift sample is 100$\%$ statistically complete down to $I_{814}$ = 
23.5, meaning that redshifts were obtained for every object targeted 
spectroscopically above that limit.  However, the redshift sample is 
not spatially complete since not all objects brighter than $I_{814}$ = 
23.5 were targeted.  Furthermore, the basic photometric catalog may 
not be complete to $I_{814}$ = 23.5 if low surface brightness galaxies 
are lost in the SExtractor selection process.  This effect is 
discussed below but does not appear to affect our major conclusions.

\subsection {Selection Effects} \label{seleff} 

Since selection effects can conceivably mimic real evolutionary 
changes in the high-redshift galaxy population, it is important to 
determine how they affect the DEEP/GSS sample.  This section describes 
selection effects both qualitatively and quantitatively.

\subsubsection{Observed Structural Parameter Distributions} 
\label{obspardist}

In order to determine whether the sample of galaxies with redshifts is 
an unbiased representation of the photometric catalog population in 
the GSS, the observed structural parameter distributions of galaxies 
with Keck redshifts were first compared to those of 725 general field 
galaxies measured in the same 6 fields of the GSS covered by the 
spectroscopic sample so far.  These galaxies were photometrically 
selected using SExtractor and then cut at $I_{814}$ = 23.5, just like 
the sample of redshift galaxies.  They should therefore be an 
equivalent sample, differing only in their lack of redshifts.  
Figure~\ref{obs-galpar} compares the distributions of half-light 
radius, bulge fraction, magnitude, color and surface brightness for 
those galaxies in the photometric catalog population (6 GSS fields) to 
those with Keck redshifts.  There are indeed no discernible 
differences between galaxies in the DEEP/GSS redshift sample and the 
photometric catalog population of the GSS fields.  The bulge fraction 
distributions look nearly identical, and the color distributions also 
look similar.  Finally, and most importantly, the surface brightness 
distributions show that the galaxies with DEEP/GSS Keck redshifts are 
an unbiased sample of the surface brightnesses of the photometric 
catalog.  A detailed analysis of the survey selection function is 
carried out in the next section.

\subsubsection{Determination of the DEEP/GSS Selection 
Function for Disk-Dominated Galaxies} \label{qselfunc}

The observed distribution of galaxies in the $M_{B_{0}}-R_{d}$ plane 
as a function of redshift $\Psi(M_{B_{0}},R_{d},z)$ is the result of 
any inherent changes in the resident\footnote{It is very important to 
note the use of the term ``resident'' here and throughout the rest of 
the paper to refer to the intrinsic galaxy population at a given 
redshift $z$.  In the absence of real evolution in the galaxy 
population with redshift, all resident populations would be the same 
as the local population of galaxies.} galaxy distribution 
$\Psi_{U}(M_{B_{0}},R_{d},z)$ in that plane and of observational 
selection effects.  Selection effects are likely to be significant 
given the wide range of sizes and surface brightnesses observed 
locally (\cite{bender92}, \cite{burst97}).  It is therefore important 
to carefully characterize selection effects to disentangle them from 
real changes in $\Psi_{U}(M_{B_{0}},R_{d},z)$.  The path from 
$\Psi_{U}(M_{B_{0}},R_{d},z)$ to $\Psi(M_{B_{0}},R_{d},z)$ is given 
by:

\begin{eqnarray}
\Psi(M_{B_{0}},R_{d},z) = S_{P
S} (M_{B_{0}},R_{d},z) S_{UP} (M_{B_{0}},R_{d},z) \nonumber \\
 \Psi_{U} (M_{B_{0}},R_{d},z), 
\label{seleq}
\end{eqnarray}

\noindent where $M_{B_{0}}$ is the rest-frame B-band absolute 
magnitude and $R_{d}$ is the disk scale length in kpc (note 
the use of lower and upper cases to distinguish between apparent and 
intrinsic radii throughout this paper).  The subscript $UP$ stands for 
``Universe to Photometric sample'', and the subscript $PS$ stands for 
``Photometric sample to Spectroscopic sample''.  The resident galaxy 
distribution $\Psi_{U}(M_{B_{0}}, R_{d},z)$ is not known {\it a 
priori}.  Once the two selection functions in Equation~\ref{seleq} 
have been characterized, their product (denoted $S_{US}$ hereafter) 
shows the region of the $M_{B_{0}}-R_{d}$ plane where real galaxies 
would have been observed if they existed in that region at high 
redshift.

The selection function $S_{UP}(M_{B_{0}},R_{d},z)$ contains the 
information needed to go from any sample of galaxies on the sky to the 
photometric catalog produced with SExtractor and reflects the adopted 
SExtractor detection parameters (detection threshold in sigmas, 
minimum detection area, etc.).  In practice, $S_{UP}(M_{B_{0}},R_{d},z)$ 
is derived from the selection function 
$S_{UP}(I_{814},r_{d})$ determined as a function of the observed 
apparent magnitude $I_{814}$ and the apparent disk scale length 
$r_{d}$ in arcseconds. The transformation $S_{UP}(I_{814},r_{d}) \rightarrow
S_{UP}(M_{B_{0}},R_{d},z)$ was made in each redshift bin using 
$k$-corrections calculated with the median observed galaxy 
$V_{606}-I_{814}$ color of the $B/T \leq 0.2$ galaxies at that redshift 
and the assumed cosmology.

$S_{UP}(I_{814},r_{d})$ was constructed by generating 50,000 $B/T = 0$ 
galaxy models with structural parameter values uniformly covering the 
ranges: $16.0$ $\leq$ $I_{814} \leq 25.0$, $0\arcsecpoint0 \leq r_{d} \leq 
10\arcsecpoint 0$, $0\deg \leq i \leq 85\deg$.  Each model galaxy was 
added, one at a time, to an empty 20\arcsec $\times$ 20\arcsec section 
of a typical $HST$ GSS image.  ``Empty'' here means that no objects 
were detected by SExtractor in that sky section with the same 
detection parameters used to construct the object catalog.  Using an 
empty section of the GSS ensured that $S_{UP}(I_{814},r_{d})$ was 
constructed with the real background noise that was seen by the 
detection algorithm.  The background noise included read-out, sky and 
the brightness fluctuations of very faint galaxies below the detection 
threshold.  This last contribution to the background noise is 
particularly hard to model theoretically, and the current approach 
bypassed this problem.  SExtractor was run on each simulation with the 
same parameters that were used to build the SExtractor catalog.  The 
function $S_{UP}(I_{814},r_{d})$ was taken to be the fraction of 
galaxies successfully detected and measured by SExtractor at each 
value of $(I_{814},r_{d})$.

The resulting $S_{UP}(I_{814},r_{d})$ is shown in the upper left-hand 
panel of Figure~\ref{obs-selecf}.  Since the SExtractor detection 
algorithm depends mostly on surface brightness, it is easy to 
understand why $S_{UP}(I_{814},r_{d})$ is bounded at low surface 
brightness by a line of nearly constant surface brightness, i.e., a 
straight line in the $M_{B_{0}}-R_{d}$ plane with slope close to $-$5.  
At the faint end, the boundary curves down as objects reach a total 
magnitude limit where they cannot meet the surface brightness 
detection threshold for any size.  The upper center panel shows the 
magnitude distribution $\Psi_{P}(I_{814})$ of the general GSS 
population in the SExtractor catalog, and the upper right-hand panel 
shows the magnitude distribution $\Psi_{S}(I_{814})$ of galaxies with 
Keck redshifts.  The lower left-hand panel of Figure~\ref{obs-selecf} 
shows the second observed selection function, $S_{PS}(I_{814},r_{d})$ 
which is needed to go from the photometric catalog $\Psi_{P}(I_{814})$ 
to the spectroscopic sample $\Psi_{S}(I_{814})$.  Even though the 
DEEP/GSS sample is statistically complete down to $I_{814} = 23.5$ 
(i.e., a redshift was obtained for every galaxy targeted 
spectroscopically), $S_{PS}(I_{814},r_{d})$ should not be expected to 
simply be a flat function down to that magnitude limit but will rather 
show a negative gradient since the fraction of galaxies in the 
photometric catalog targeted spectroscopically decreases with 
increasing magnitude.  $S_{PS}(I_{814},r_{d})$ is computed in the 
observed $I_{814}-r_{d}$ plane according to the rules described in 
Appendix~\ref{appA} and then transformed to the rest frame 
$S_{PS}(M_{B_{0}},R_{d},z)$ in each redshift bin with the same 
$k$-corrections as for $S_{UP}(I_{814},r_{d})$.  Note that the cutoff 
at $I_{814} = 23.5$ has been applied.  $S_{PS}(I_{814},r_{d})$ is flat 
at $I_{814} \leq 19.0$ since any galaxy above that magnitude limit 
would have been targeted spectroscopically.

The final selection function $S_{US}(I_{814}$, $r_{d}$), which is the 
product of $S_{UP}$ and $S_{PS}$, is shown in the lower right-hand 
panel of Figure~\ref{obs-selecf}.  The apparent sizes and magnitudes 
of galaxies with Keck redshifts above $I_{814} = 23.5$ are also 
plotted over the filled contours of $S_{US}$.  Note that the observed 
density of points does not necessarily follow the contours of $S_{US}$ 
since bright galaxies, though they have a greater probability of being 
detected, are also less numerous in the Universe.  The rest-frame 
counterparts of $S_{US}(I_{814}$, log $r_{d}$) at each redshift are 
presented in the next section in conjunction with the magnitude-size 
relation for disk galaxies as a function of redshift.

\section{The Magnitude-Size Relation} \label{sizemag}

\subsection{Disk galaxies} \label{sizemag_disk}

Figure~\ref{sizemag-disk} shows the relation between intrinsic disk 
scale length and disk rest-frame absolute B-band magnitude as a 
function of redshift for galaxies with $B/T < 0.5$ in the DEEP/GSS 
sample.  The redshift ranges of the lowest three bins are nearly 
identical to those of Schade et al.  (1996a,b).  However, the Keck 
sample goes 1.5--2.0 magnitudes deeper than their CNOC sample and also 
provides two bins at higher redshifts between $z = 0.70$ and $z = 
1.10$.  The error bars on the points are the $99\%$ confidence 
intervals of the $I_{814}$ and $r_{d}$ measurements.  The open 
triangles are objects with structural parameters measured with the 
same software as the rest of the objects but whose redshifts come from 
the CFRS survey.  The redshift range covered by all five bins 
corresponds to a total lookback time interval of 5.8 Gyrs 
($q_{0}=0.1$).

The long-dashed line is the Freeman relation for an exponential disk 
B-band central surface brightness of 21.65 mag arcsec$^{-2}$ 
(\cite{freeman70}).  The disk sizes and luminosities of most galaxies 
in the two lowest redshift bins cluster near this relation.  
Galaxies in the third redshift bin also cluster around the Freeman 
relation, although a substantial number of objects begin to exhibit 
systematically higher central surface brightness than the Freeman 
value.  Major deviations from the Freeman relation are seen above $z = 
0.70$, with evidence for a class of very small, high-surface brightness 
objects at $z \ge 0.90$.  It is interesting to observe that a 
significant number of galaxies remain on the Freeman locus even at $z 
\ge 0.90$, consistent with the existence of a population of Freeman 
surface brightness disks even at this early epoch.  The large spread in 
surface brightness at $z \ge 0.90$ may be due either to a strong 
differential disk evolution that leaves some galaxies unaffected or to 
a new galaxy population making its entrance at $z \ge 0.90$.  Both 
possibilities are discussed further in Section~\ref{discuss}.

The magnitude-size relations shown in Figure~\ref{sizemag-disk} cannot 
be properly interpreted without taking selection effects into account.  
In Figure~\ref{rest-selecf}, the rest-frame selection function 
$S_{US}$ for each redshift bin is shown as the shaded contours, and 
the same galaxies as in Figure~\ref{sizemag-disk} are replotted to 
compare their location to the predictions of the selection function.  
The Freeman disk relation is also replotted for reference.  Two 
important observations can be made from Figure~\ref{rest-selecf}.  
First, very few disks are observed at low redshifts in the ranges $-20 
\le M_{B_{0}}(disk) \le -18$ and $0.5 \le$ log $R_{d} \le 1.2$ 
(significantly above the Freeman relation) even though they are 
detectable in principle. Second, the envelope of the selection 
function in the two highest redshift bins comes close to the Freeman 
locus, and this situation has a very serious implication: galaxies 
whose surface brightness may have been {\it fainter} at those 
redshifts are {\it not} detectable in the DEEP/GSS sample.  Hence, 
this sample cannot be used to distinguish between a scenario in which 
mean surface brightness is higher at high redshift and one in which 
mean surface brightness remains unchanged but the {\it spread} in 
surface brightness about that mean value increases.

Although the present sample may be somewhat biased to high surface 
brightness at high redshift, this effect would appear to be even more 
pronounced in the Schade et al.  (1996a,b) and Roche et al.  (1998) 
samples.  This is shown in Figure~\ref{compare-loci}, which 
illustrates loci in the magnitude-size plane occupied by the various 
samples as a function of redshift.  The DEEP/GSS loci are taken from 
Figure~\ref{sizemag-disk} and are shown by the solid boxes.  Loci for 
the other samples have been drawn by eye from analogous plots in the 
published papers.  These boxes cannot substitute for a quantitative 
treatment, but even this rough comparison is instructive.

Systematically, it is seen that the loci of the Schade et al.  and 
Roche et al.  samples lie at higher surface brightness relative to the 
DEEP/GSS sample, and that this offset increases with redshift.  Put 
differently, the first two samples fail to include many low surface 
brightness objects of the sort actually detected by DEEP/GSS. 
Quantitatively, at $z \sim 0.7$, the upper boundary difference 
between the first two samples and DEEP/GSS is about 0.3 dex in log 
$R_{d}$.  Loss of these galaxies would translate to a shift of about 
0.75 mag in mean surface brightness if the DEEP/GSS box were uniformly 
populated vertically -- about the same magnitude as the surface 
brightness evolution actually found in the other two samples.  
Figure~\ref{compare-loci} thus raises the possibility that the surface 
brightness evolution seen by Schade et al.  and Roche et al.  might 
have been caused in large part by the loss of low surface brightness 
objects at high $z$.

Figure~\ref{compare-loci} also shows boxes marking the distant compact 
galaxy sample of Phillips et al.  (1997).  This sample represents the 
top 35$\%$ of galaxies in surface brightness to $I_{814} = 23.5$ 
around the Hubble Deep Field.  The location of the Phillips et al.  
sample in Figure~\ref{compare-loci} is consistent with its compact 
high surface brightness nature.  Further comparison is made to this 
sample below.

As a first step to study in greater detail whether the magnitude-size 
relations shown in Figure~\ref{sizemag-disk} exhibit real evolution or 
not, surface brightness distributions were calculated for each 
redshift bin.  The disk rest-frame B-band central surface brightness 
of each galaxy was computed using the relation:
\begin{equation}
	 \mu_{0_{B}} = M_{B_{0}} + 5.0\thinspace {\rm log} R_{d} + 38.57,
	\label{mu0}
\end{equation}
\noindent where $M_{B_{0}}$ is the disk rest-frame B-band absolute 
magnitude, and $R_{d}$ is the disk scale length in kpc.  
Equation~\ref{mu0} is valid for pure exponential disks, and the units 
of $\mu_{0_{B}}$ are mag arcsec$^{-2}$.  The resultant surface 
brightness distributions shown in Figure~\ref{rest-sb-histo-nosel} are 
raw without correction for selection effects, and the area under each 
distribution is normalized to unity.  The mean of each distribution is 
shown as a large filled circle, and the error bar attached to each 
circle is 3 times the standard error of the mean.  The vertical dotted 
lines mark the canonical Freeman disk central surface brightness of 
21.65 mag arcsec$^{-2}$.  There is a systematic increase in the mean 
disk central surface brightness from $\mu_{0_{B}}$ = 21.7 mag 
arcsec$^{-2}$ at $z=0.20$ to $\mu_{0_{B}}$ = 20.4 mag arcsec$^{-2}$ at 
$z=0.80$.  This raw increase of 1.3 mag is similar to the effect seen 
by Schade et al.  and Roche et al.  over the same redshift range.

Next, the surface brightness distributions were re-calculated by 
taking the survey selection function into account.  Let 
$\phi_{ij}(\mu_{0_{B_{ij}}})d\mu_{0_{B_{ij}}}$ be the number of 
galaxies in the $j^{th}$ bin of the surface brightness distribution 
for the $i^{th}$ redshift bin.  Adjusted surface brightness 
distributions in Figure~\ref{rest-sb-histo-sel} were calculated using 
the equation:

\begin{equation}
      \phi_{ij}(\mu_{0_{B_{ij}}}) d\mu_{0_{B_{ij}}} = \sum_{k} 
      {{S_{US}(R_{d_{ijk}}, M_{ijk}, 0.90 \leq 
      z)}\over{S_{US}(R_{d_{ijk}}, 
      M_{ijk}, z_{i})}}
      \label{mu0dist}
\end{equation}

\noindent where the index $k$ runs over all the galaxies in the 
$i^{th}$ redshift bin with surface brightness between 
$\mu_{0_{B_{ij}}}$ and $\mu_{0_{B_{ij}}} + d\mu_{0_{B_{ij}}}$.  Each 
object was thus de-weighted by the inverse ratio of the value of the 
selection function in its redshift bin and magnitude-size position to 
the value of the selection function in the highest redshift bin at the 
same magnitude-size position.  This weighting scheme effectively 
applies the selection function of the highest redshift bin uniformly 
to all the redshift bins.  The area under each distribution in 
Figure~\ref{rest-sb-histo-sel} was again normalized to unity.  The 
parameters of the surface brightness distributions with and without 
selection function corrections are given in 
Table~\ref{sb_histo_param}.  These distributions show that there is 
{\it no} detectable systematic change in the mean disk central surface 
brightness over the redshift range $0.1-1.1$ when a uniform selection 
function is applied to all bins.  Why has this occurred?  Previous 
discussion hinted that loss of low surface brightness galaxies at high 
$z$ might be important.  However, the major consequence of applying 
the highest redshift selection function to all bins is actually the 
deletion of {\it intrinsically faint} galaxies below $M_{B_{0}} = -19$ 
mag from the nearer redshift bins.  Inspection of 
Figure~\ref{rest-selecf} shows that the remaining, brighter galaxies 
tend to lie {\it below} the Freeman relation, i.e., at higher 
surface brightness, at all redshifts.  In other words, once comparison 
is restricted to galaxies of similar (i.e., bright) absolute 
magnitude, no significant evidence for evolution in mean disk surface 
brightness remains, at least for galaxies in the range $M_{B_{0}} 
(disk) = -19$ to $-22$ mag.

\subsection{Global Population} \label{sizemag_all}
 
The analysis so far has been restricted to the disks of disk-dominated 
galaxies with $B/T < 0.5$.  To extend the analysis to the whole 
distant galaxy population, a local magnitude-size relation must be 
chosen that covers as many local galaxies as possible.  The choice of 
a local magnitude-size relation is clearly key to interpreting the 
relation at higher redshifts.  Schade et al.  (1996a,b) used fits to 
galaxies in the cluster Abell 2256 ($z = 0.06$) as a local reference.  
Roche et al.  (1998) used a set of four galaxy type-dependent 
relations (Freeman, E/S0, Sab-Sbc, and Scd-Im).  While these relations 
are useful, they do not represent very well the wide range of local 
galaxy photometric parameters over all Hubble types.

To broaden the approach, half-light radii $R_{hl}$ are introduced as a 
measure of galaxy size.  Half-light radii are available for all local 
Hubble types in quantity in the RC3 catalog.  Specifically, the sample 
of 957 local galaxies extracted from the RC3 by Burstein et al.  
(1997; BBFN) is used here.  Figure~\ref{local-kappa} shows half-light 
radius versus $M_{B_{0}}$ for this sample.  The data are divided into 
four broad morphological classes: ellipticals (E), early-type spirals 
(Sa-Sbc), late-type spirals (Sc-Sdm), and irregular galaxies (Sm-Irr).  
The Freeman relation is again shown as a reference (now expressed 
using half-light radius).  This large dataset demonstrates how 
significant the total spread in the local magnitude-size relations is: 
$\sim 1.5$ mag for the bright ellipticals and $\sim 3.0$ mag for the 
late-type spirals.  More significant is the fact that the different 
local Hubble types differ systematically in radius (and, therefore, in 
surface brightness) at a given absolute magnitude.  
Figure~\ref{local-kappa} shows that, at $M_{B_{0}} = -22$ mag, bright 
Sa-Sc spirals are nearly 2 magnitudes brighter in mean surface brightness 
than ellipticals; however, at $M_{B_{0}} = -18$ mag, they are 2 
magnitudes dimmer; Sm and Irr galaxies at $M_{B_{0}} = -18$ are a 
further 0.5 magnitude dimmer yet than Sc's.

These systematic effects are of crucial importance in predicting what 
distant surveys will show. Suppose, for the sake of argument, that the 
distant galaxy population is the same as that seen locally. The 
distant selection function will then cut through the manifold of 
galaxies in different locations at different redshifts, turning up 
galaxies in proportion to their resident number densities at each 
redshift. At nearby redshifts, the selection function will turn up 
dim but abundant Sm and Irr types. At distant redshifts, it will turn 
up a mixture of rarer but brighter E's and early-type spirals. Mean 
surface brightness averaged over all detected galaxies will thus tend 
to ``walk'' from low to high values with distance, precisely as seen 
in the raw data in Figure~\ref{rest-selecf}. (It may even walk at 
fixed absolute magnitude, depending on the relative volume densities 
of different Hubble types and how their surface brightnesses differ 
at that particular magnitude.)

Finally, it should be stressed that the BBFN catalog in 
Figure~\ref{local-kappa} is {\it not} a volume-limited sample, nor 
does it fairly sample the full range in surface brightnesses seen 
locally, from high surface brightness starbursting HII and compact 
narrow emission-line galaxies on the one hand, to low surface 
brightness dwarf spheroidals on the other. Comparisons must therefore 
be made with caution, taking note of these deficiencies. These issues are 
discussed further below.

Measured half-light radii of all DEEP/GSS galaxies are 
plotted against total rest-frame B-band absolute magnitudes in 
Figure~\ref{sizemag-all}.  Filled circles are disk-dominated galaxies 
($B/T < 0.5$), and open circles are bulge-dominated galaxies.  The 
solid line is the Freeman relation.  Figure~\ref{sizemag-all} again 
shows that a significant number of galaxies remain on or close to the 
Freeman relation at all redshifts, and further that such galaxies are 
largely disk-dominated.  Bulge-dominated galaxies appear to lie at 
higher surface brightnesses than disk-dominated galaxies, especially 
at the highest redshifts.

To facilitate further comparison, Figure~\ref{sizemag-local-distant} 
directly overplots the distant half-light radii from 
Figure~\ref{sizemag-all} on top of the total local catalog from 
Figure~\ref{local-kappa} (tiny dots). As expected, at low redshift, 
the DEEP/GSS sample is dominated  by intrinsically faint but abundant 
late-type galaxies deep in the luminosity function. These galaxies are 
largely disk-dominated, with $B/T < 0.5$ (solid dots, 
Figure~\ref{sizemag-all}) and occupy the same magnitude-size 
locus as local Sm-Irr galaxies (Figure~\ref{local-kappa}), which also 
have pure exponential profiles (\cite{gallagher84}). The excellent 
agreement between the two loci is not surprising given the small 
difference in epochs between the two samples.

As redshift increases, the observed magnitude-size envelope begins to 
shift, both in surface brightness and also in absolute magnitude. 
However, out to $z = 0.70$, the projected distributions in the 
magnitude-size plane remain roughly consistent with those predicted by 
the local envelope. There may be a slight shift to higher surface 
brightness at $M_{B_{0}} \lta -20$ in the $z$ = 0.50--0.70 bin, but 
the number of points is small. Overall, the impression is one of no 
great change from local galaxies out to $z = 0.70$. 

Beyond $z$ = 0.70, there is a hint of change.  First, the range of 
surface brightnesses in the $z$ = 0.70--0.90 bin appears to be larger 
than that seen locally, and a handful of very high surface brightness 
objects (HSBs hereafter) has appeared.  The latter effect cannot be 
due to observational errors on the radii, as they are quite small (cf.  
Figure~\ref{sizemag-disk}).  Denoted by the open circles in 
Figure~\ref{sizemag-local-distant}, these HSBs increase still further 
in number in the $z$ = 0.90--1.10 bin, where objects distinctly more 
compact than the local BBFN sample are seen.  Remarkably, HSBs span 
the whole range of bulge fractions and rest-frame colors and therefore 
may not be restricted to a single structural type.

HSBs aside, the bulk of galaxies at $z > 0.7$ are notable more for 
their resemblance to the local sample than for their differences.  
This statement is tentative, as no quantitative comparison has yet 
been made between the near and distant samples taking {\it resident 
local densities} of the various structural types properly into 
account.  It is also conceivable that further differences will emerge 
once velocity widths are measured (a survey of these is in progress).  
For now, the present results suggest simply that no great global 
evolution has occurred in the surface brightnesses of luminous 
galaxies since $z \sim 1$.  This conclusion is based on a considerably 
deeper and more complete sample than heretofore available, as 
Figure~\ref{compare-loci} shows.

Strictly speaking, one should also carry out selection function 
experiments on half-light radii of the various Hubble types, analogous 
to those carried out for disk radii in Figure~\ref{rest-selecf}. 
However, the main effect of these experiments for bright galaxies was 
to limit the sample in magnitude, not in surface brightness. Thus, 
neither the relative constancy in surface brightness nor the 
appearance of luminous high surface brightness galaxies are likely to 
stem from as-yet-uncalibrated selection effects.

\section{Discussion} \label{discuss}

The inclusion of selection effects in the analysis of the disk galaxy 
magnitude-size relation at high redshift has led to a number of 
important results.  

First, even though the DEEP/GSS survey goes nearly two magnitudes 
deeper than previous studies, the faint envelope of the survey 
selection function in surface brightness approaches the local Freeman 
relation at redshifts greater than $z = 0.70$.  This means that the 
magnitude-size relation of past and present samples cannot be used by 
itself to favor a scenario in which the mean surface brightness of 
disks brightens systematically with redshift over a scenario in which 
the spread in disk surface brightness increases.  Although theoretical 
expectations for the characteristic surface brightness of ``normal'' 
bright disk galaxies are comparable to the canonical central Freeman 
value of 21.7 $\pm$ 0.3 mag arcsec$^{-2}$, the spread about that value 
could be as large as $\pm$ 3.4 mag depending on the range of disk 
masses and mass-to-light ratios (\cite{dalc97}).  The disks on the 
faint end of that surface brightness distribution would quickly be 
excluded from the DEEP/GSS and other samples at relatively low 
redshifts.

Second, selection effects produce an apparent systematic increase in 
disk mean surface brightness that, at first sight, looks like luminosity 
and/or size evolution (Figure~\ref{rest-sb-histo-nosel}).  However, 
there is no such systematic increase below $z = 1.1$ when the selection 
function is used to weigh the disk surface brightness distributions as 
a function of redshift.  According to the selection function $S_{US}$, 
disks with absolute magnitudes brighter than $M_{B_{0}} = -19$ remain 
detectable over all redshift bins, and these are the disks that 
effectively contribute to the weighted surface brightness 
distributions.  Significant evolution in the surface brightness of 
such disks is not apparent, though the numbers are still small.  
Selection effects are too important for fainter disks to determine 
whether or not their surface brightness is evolving with redshift.

Third, it is apparent from Figures~\ref{sizemag-disk} 
and~\ref{sizemag-all} that a significant number of galaxies remain 
close to the canonical Freeman relation at all redshifts.  This 
population may be the same as that detected in the Tully-Fisher 
zeropoint studies of distant galaxies by Vogt et al.  (1996,1997).  
The Vogt galaxies were selected to include the largest (i.e., best 
resolved) morphologically normal spirals at each apparent magnitude, a 
criterion that favors lower surface brightness galaxies in each 
redshift bin.  It is probable that the Vogt et al.  studies also 
targeted late-type spirals preferentially, since they have stronger 
emission lines.  The constancy of the TF zeropoint for these 
particular objects would then not be surprising, since the 
star-formation timescales of late-type galaxies are quite long 
($\tau_{SFR} \ge 7$ Gyrs, \cite{bruzual93}), and mass-to-light ratios 
for such stellar populations evolve only slightly.

These long timescales are a feature of the Pure Luminosity Evolution 
model of Roche et al.  (1998), which predicts that the surface 
brightness of late-type spirals should remain constant out to $z = 1$.  
On the other hand, their Size-Luminosity Evolution model 
predicts that the same late-type galaxies should increase nearly 1.5 
magnitudes in surface brightness to the same redshift (Figure 7, 
\cite{roche98}).  This latter model does not appear to be consistent 
with the DEEP/GSS data.  Indeed, according to 
Figure~\ref{sizemag-all}, such a large brightening, if present in all 
disk galaxies, would considerably decrease the number of galaxies on 
the Freeman relation.  This deficiency of galaxies on the Freeman 
relation could not be compensated by increasing the surface brightness 
of any resident Irr population, even if present.  The internal 
kinematics of Irr galaxies cannot reproduce the observations of Vogt 
et al.  The most massive Irr galaxies have rotation velocities lower 
than 100 km s$^{-1}$, and typical Irr galaxies have $V_{c} \sim  
50 - 70$ km s$^{-1}$ (\cite{gallagher84}).

As noted earlier, the last redshift bin ($z \geq 0.9$) contains about 
nine luminous, high surface brightness (HSB) objects (shown as open 
circles in Figure~\ref{sizemag-local-distant}) that occupy a region of 
the magnitude-size plane rarely populated by local galaxies.  The 
exact number of these aberrant objects depends on the boundary adopted 
for the envelope of normal local galaxies; a conservative, high 
surface brightness boundary was adopted here, minimizing the number of 
HSBs.  The nature of the HSBs will be the subject of a future paper.  
For now, it is sufficient to note that the present HSBs occupy 
essentially the same region of the magnitude-size plane as the distant 
compact sample of Phillips et al.  (1997), as shown in 
Figure~\ref{compare-loci}.  Like the present HSBs, that sample 
exhibited a mixture of colors and emission-line strengths, and many 
also had structural parameters similar to the distant compact narrow 
emission-line galaxies of Guzm\'an et al.  (1998).  Based on colors 
and spectra, Phillips et al.  concluded that their population of HSBs 
was rather heterogeneous, though with a tendency to low internal 
velocity widths compared to local galaxies of similar absolute 
magnitude.  The present data give further support to the notion that 
HSBs are a mixed population.  Furthermore, it is also worth noting 
that the magnitude-size locus of the $z \sim 3$ galaxies observed by 
Lowenthal et al.  (1997) overlaps with some HSBs (see dashed box in 
Figure~\ref{sizemag-local-distant}).  Although this overlap raises 
interesting possibilities, it is by itself insufficient to directly 
link some HSBs to $z \sim 3$ galaxies. Potential links between these 
two types of galaxies will be explored in a future paper using a 
full set of structural parameters and emission linewidths.

At least three explanations for HSBs suggest themselves: (1) They may 
represent an entirely new population of galaxies not found locally and 
making its appearance at $z \sim 1$.  (2) The local magnitude-size 
dataset used as a comparison here may be incomplete.  Indeed, BBFN 
warn that rare types of local HSB galaxies such as starburst HII galaxies 
may be missing from their sample.  The blue colors of some HSBs 
strongly suggest that at least some are starbursts.  (3) HSB galaxies 
may have had high surface brightness in the past but have faded to 
become a normal population locally.  For example, at a fixed radius, 
most HSB galaxies would have to fade by at least 1 mag to fall within 
the local magnitude-size envelope today.  This is consistent with the 
passive evolution of a very early-type population, and it would nicely 
explain red HSBs.  In view of the heterogeneous nature of HSBs, all 
three scenarios may be at work.

\section{Conclusions} \label{conc}
The $HST$ structural parameters of 190 galaxies ($I_{814} \leq 23.5$) 
in the Groth Survey Strip with DEEP Keck redshifts have been used to 
build a magnitude-size relation in five redshift bins over the range 
0.10 $ \le z \le 1.10$.  Selection functions were quantitatively 
determined for all five redshift bins.  The major conclusions are: 
\vskip 12 true pt

1.  If selection effects are ignored, the mean surface brightness of 
disks in the sample appears to increase systematically by $\sim$ 1.3 
mag from $z = 0.1$ to $z = 0.9$.  Such an increase was also found by 
Schade et al.  (1996b) and Roche et al.  (1998).

2.  However, if the survey selection functions are used to correct the 
surface brightness distributions in the different redshift bins, then 
there is {\it no} detectable change out to $z = 0.9$ in the mean 
surface brightness of disks brighter than $M_{B_{0}} = -19$.  These 
disks are the faintest objects not significantly affected by selection 
effects in the highest redshift bin.

3.  A number of disks remain on the canonical Freeman relation in all 
redshift bins.  This is consistent with the lack of evolution seen in 
the zeropoint of the high-redshift Tully-Fisher relation of Vogt et 
al., and it does not support a 1.5 mag brightening in surface 
brightness predicted for {\it all} late-type galaxies by the 
Size-Luminosity Evolution model of Roche et al.  (1998).  On the 
other hand, the constant surface brightness of late-type galaxies 
predicted by their Pure Luminosity Evolution model agrees with the 
DEEP/GSS observations of these disks.

4.  At all redshifts, two important factors conspire to produce an 
apparent increase in global galaxy surface brightness out to $z = 
0.9$: the relative volume densities of different Hubble types and the 
survey selection function.  Lower surface brightness late-type and 
irregular galaxies will dominate lower redshift bins due to their 
higher volume densities.  Higher surface brightness early-type 
galaxies will dominate the highest redshift bins as a result of 
increasing survey volume and the selection function cutting different 
slices through the resident galaxy surface brightness distribution at 
that redshift.  So, for example, even if the distant galaxies were 
identical to those seen locally, average galaxy surface brightness 
would still shift to higher values at higher redshifts, in qualitative 
agreement with what is seen.  Systematic shifts in galaxy surface 
brightness versus redshift thus may not necessarily represent real 
evolution in the radii or luminosities of the resident galaxy 
population.

5.  Nine galaxies at redshifts $z \gta 0.9$ occupy a region of the 
magnitude-size plane rarely populated by local galaxies.  As will be 
shown in a forthcoming paper, these luminous, high surface brightness 
galaxies exhibit a wide range of rest-frame colors and bulge 
fractions.  This suggests that a variety of scenarios might be 
required to explain their origin: a new population at $z \sim 1$ not 
found locally, a rare population missed by local surveys, and/or a 
familiar local population that has faded by at least 1 mag from $z 
\sim 1$ to the present epoch.

Further work is needed to fully understand the magnitude-size relation 
of the global galaxy population out to $z \sim 1$.  Prime requirements 
are a truly volume-limited sample of local galaxies plus comoving 
volume densities for the distant samples.  The size of the distant 
sample must also be increased to study the statistical properties of 
the HSB population and determine its origin.  Finally, linewidths must 
be obtained to provide full structural information, including masses.  
These additional parts of the global structural puzzle will be tackled 
in future papers.

\acknowledgments

Caryl Gronwall kindly provided the model SEDs and the code used for 
the $k-$corrections.  This work was funded by NSF grants AST 91-20005 
and AST95-29098 and NASA grants AR-06337.08-94A, AR-06337.21-94A, 
AR-06402.01-95A, AR-07531.01-96A and AR-07532.01-96A. L. S. gratefully 
acknowledges financial support from a Postdoctoral Fellowship from the 
Natural Sciences and Engineering Research Council of Canada.  David 
Burstein kindly provided an electronic copy of the BBFN database.  
Thanks go to J. Cohen and B. Oke for building the LRIS spectrograph.  
Many thanks also go to the Keck telescope operators and staff for 
their dedication and hard work in the harsh environment of the Mauna 
Kea summit.

\clearpage 

\figcaption[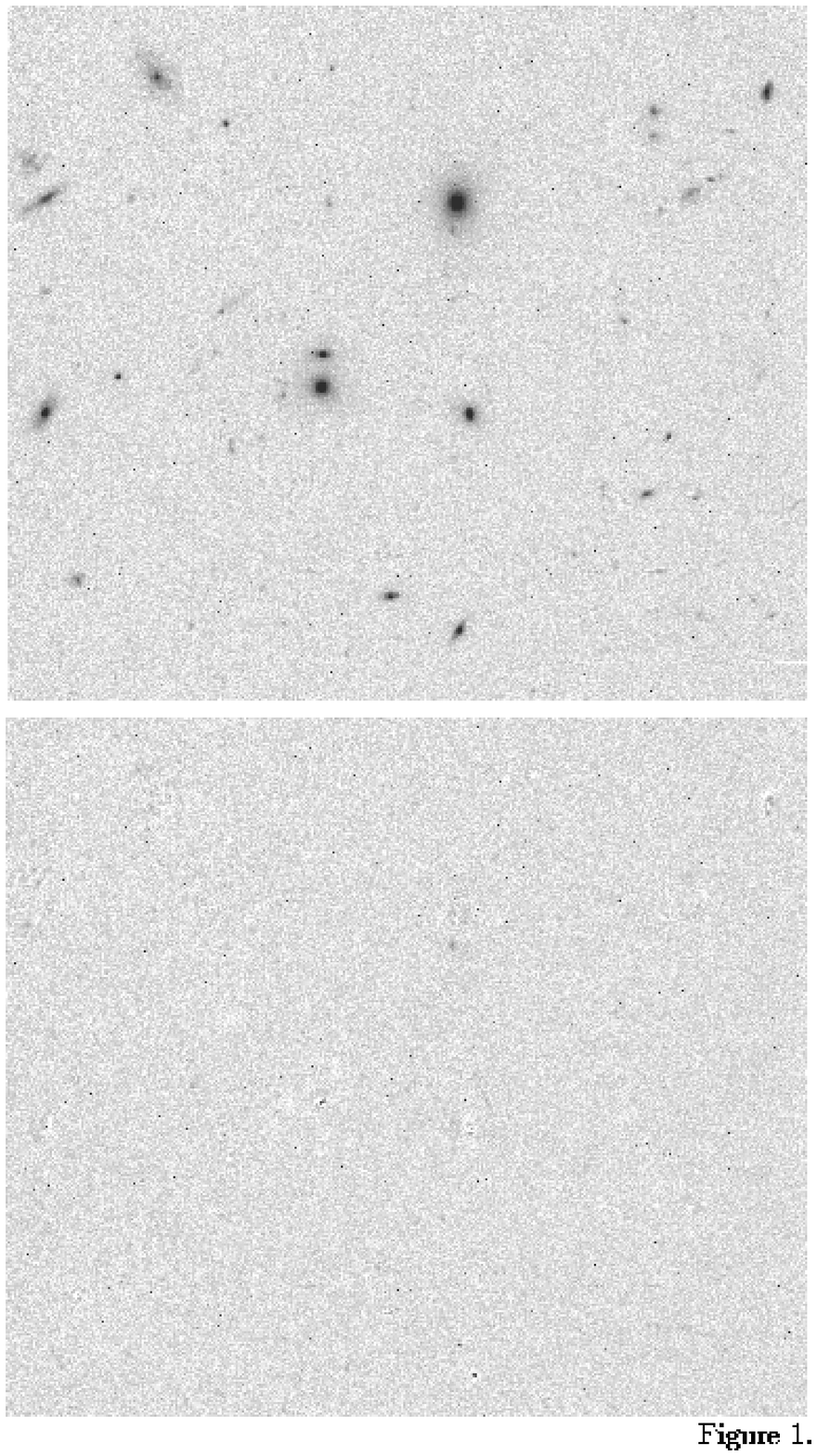]{{\it Upper Panel:} 55\arcsec $\times$ 
45\arcsec section of a typical WFPC2/F814W GSS image before GIM2D 
processing.  Exposure time was 4400s.  Galaxies in this image 
section span a wide range of bulge fractions.  {\it Lower Panel:} 
GIM2D residual galaxy image of the same image section as shown above.  
GIM2D produced this residual image by subtracting the best bulge+disk 
surface brightness model from each galaxy detected by SExtractor.
\label{gim2d-orig-res}}

\figcaption[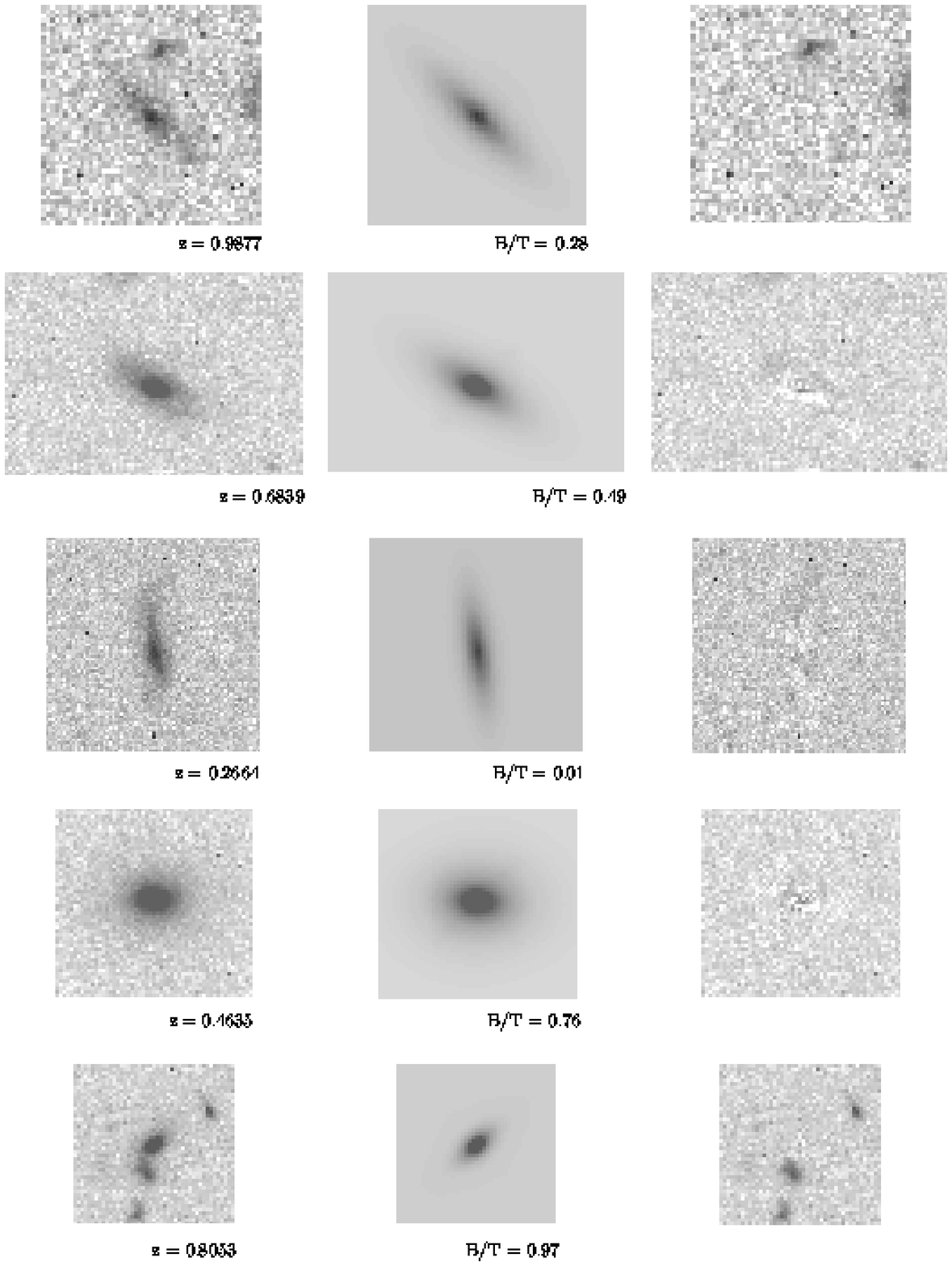]{Examples of GIM2D surface brightness fits for 
a set of five galaxies covering a range of bulge fractions and 
redshifts.  The left-hand panels are the original F814W galaxy postage 
stamp images analyzed by GIM2D, the central panels are the 
PSF-convolved best fitting surface brightness models, and the 
right-hand panels are the residual images.  Redshift and measured 
bulge fraction are given for each object.  These five galaxies have 19.9 $ 
\leq I_{814} \leq $ 22.3.
\label{gim2d-fits-examples}}

\figcaption[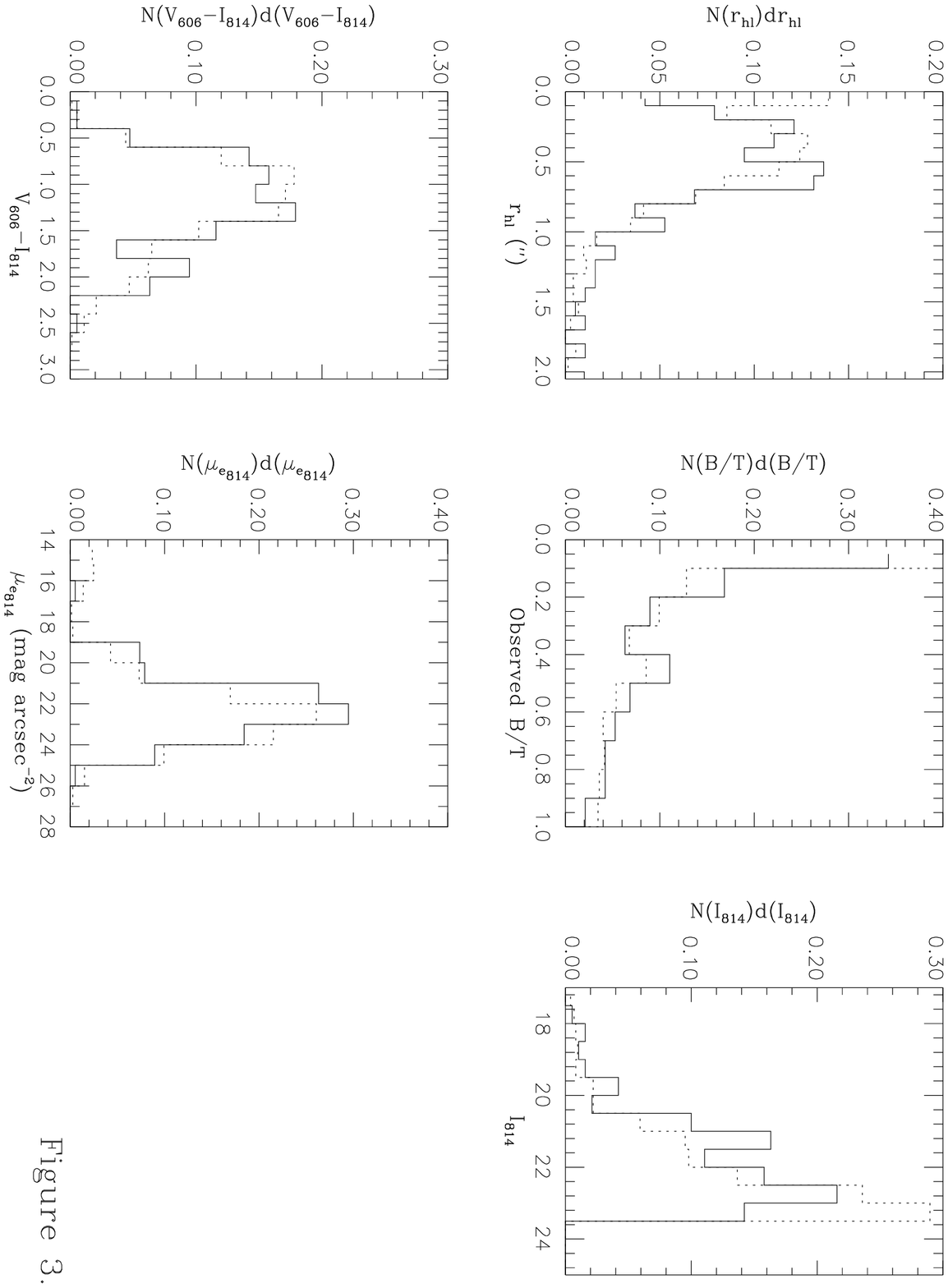]{Observed structural parameter distributions 
for galaxies with DEEP Keck redshifts ({\it solid line}) and for an 
identically selected sample of galaxies in the general GSS population 
({\it dotted line}).  From left to right and from top to bottom: 
PSF-deconvolved half-light radius $r_{hl}$, observed $F814W$ bulge 
fraction $B/T$, $F814W$ total magnitude $I_{814}$, total 
$V_{606}-I_{814}$ galaxy color, observed $F814W$ effective surface 
brightness $\mu_{e_{814}}$ in mag arcsec$^{-2}$.  Magnitudes and 
colors are in the Vega zeropoint system.  The spectroscopic sample 
shares the same characteristics as the general field population.
\label{obs-galpar}}

\figcaption[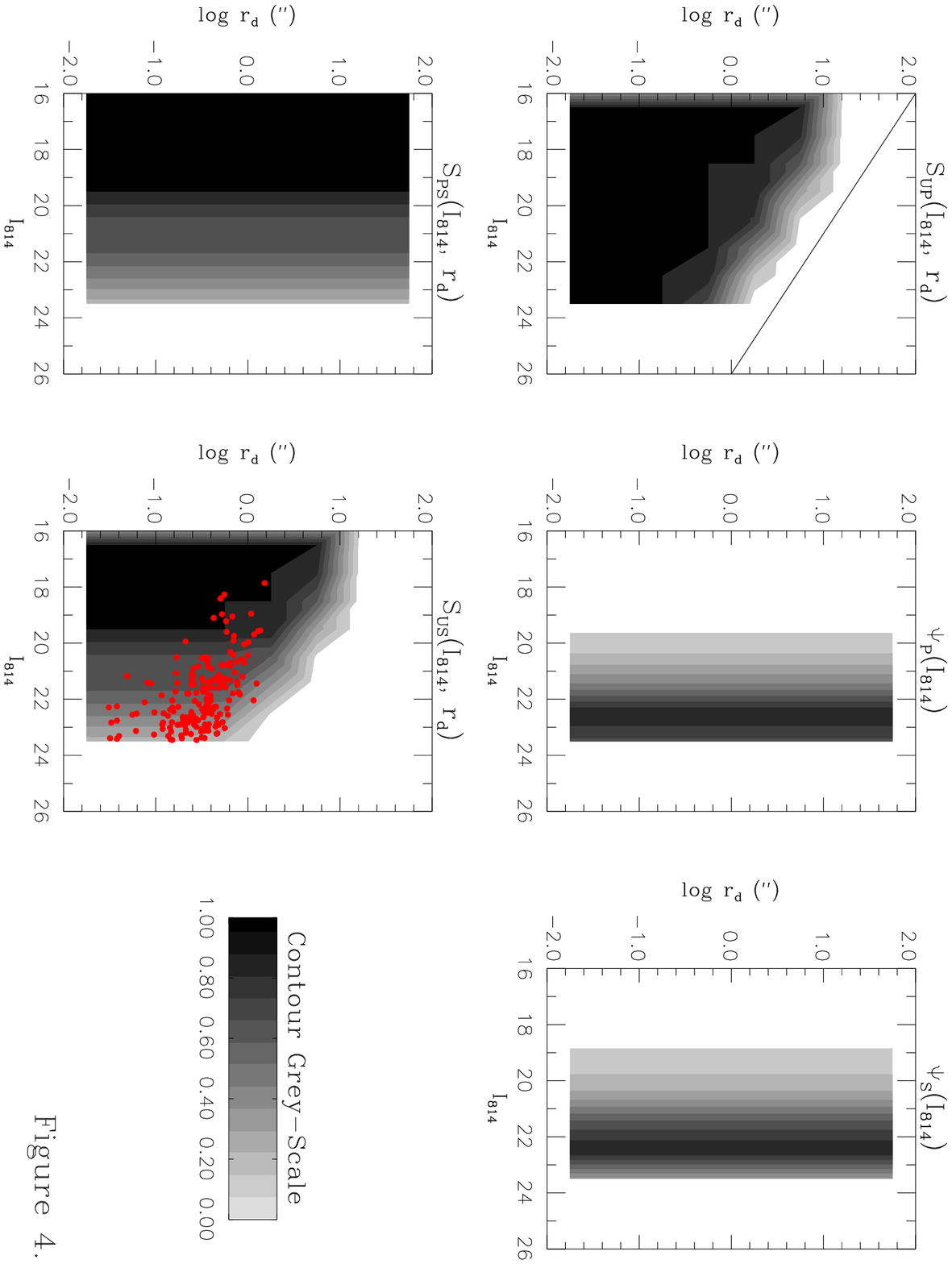]{{\it From left to right and from top to 
bottom:} SExtractor selection function $S_{UP}(I_{814},r_{d})$, 
magnitude distribution $\Psi_{P}(I_{814})$ of galaxies in the 
photometric catalog, magnitude distribution $\Psi_{S}(I_{814})$ of 
galaxies in the spectroscopic catalog, Keck spectroscopic selection 
$S_{PS}(I_{814}, r_{d})$, and combined survey selection function 
$S_{US}(I_{814}, r_{d})$.  The highest contour was normalized to one 
in each case.  The red points in the bottom right-hand panel are the 
galaxies with Keck redshifts in the sample.  The observed distribution 
of galaxies in the $I_{814}-r_{d}$ plane is given by the resident 
galaxy distribution in that plane multiplied by $S_{US}(I_{814}, 
r_{d})$.  The solid line in the upper left-hand panel is an arbitrary 
line of constant apparent surface brightness.
\label{obs-selecf}}

\figcaption[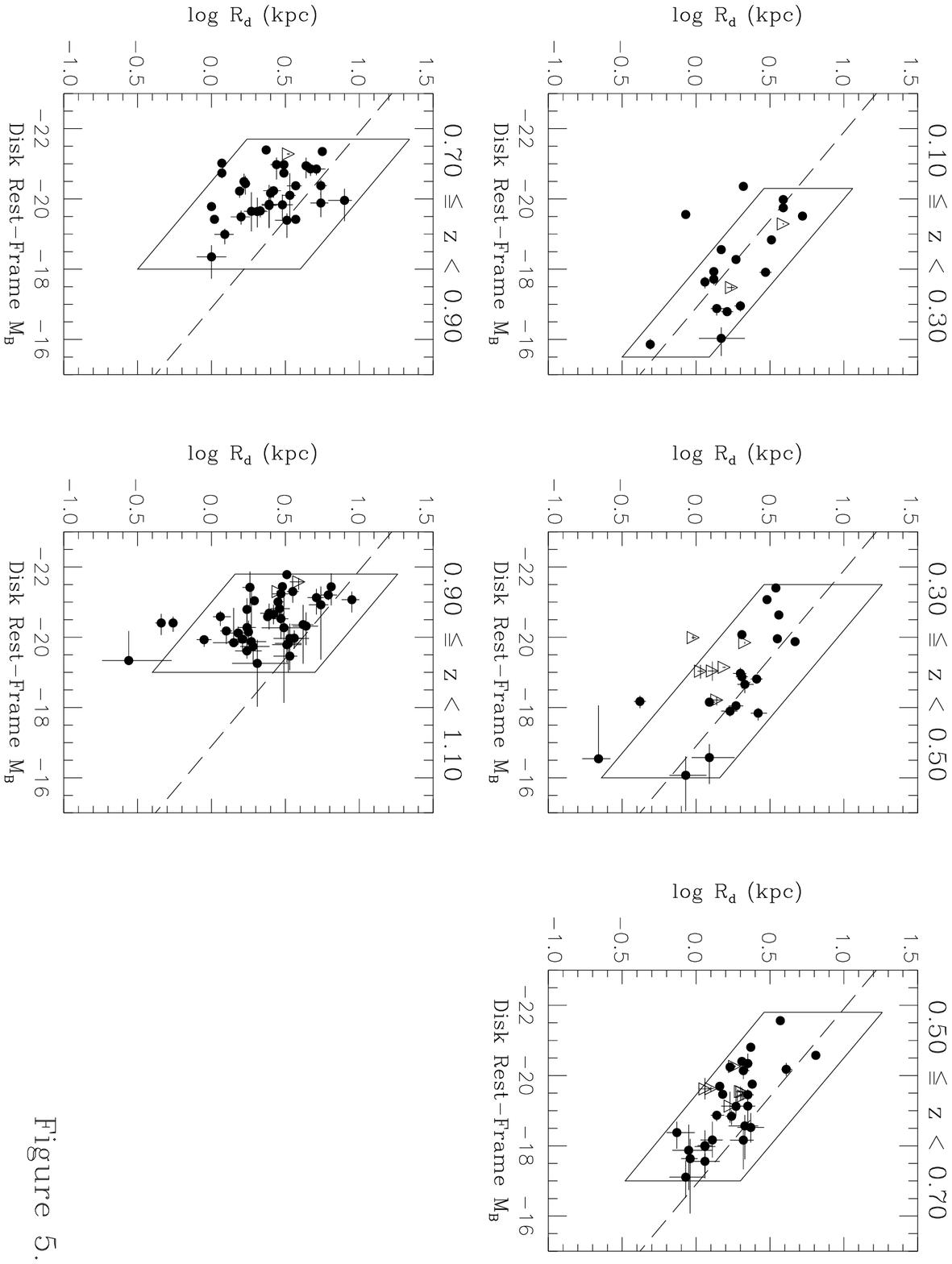]{Disk scale length versus disk rest-frame 
B-band absolute magnitude as a function of redshift for galaxies with 
$B/T < 0.5$.  {\it Filled circles:} Galaxies with Keck redshifts; {\it 
Open triangles:} Galaxies with CFRS redshifts.  {\it Long-dashed 
line:} Canonical Freeman relation for a disk central surface 
brightness $\mu_{0_{B}} = 21.65$ mag arcsec$^{-2}$ (\cite{freeman70}).  
Error bars are 99$\%$ confidence intervals.  The lowest three redshift 
bins cover nearly the same redshift ranges as those in Schade et al.  
(1996b).  The loci of the DEEP/GSS galaxies ({\it solid outlines}) are used 
later in the comparison with other samples (Section~\ref{sizemag_disk} 
and Figure~\ref{compare-loci}).
\label{sizemag-disk}}

\figcaption[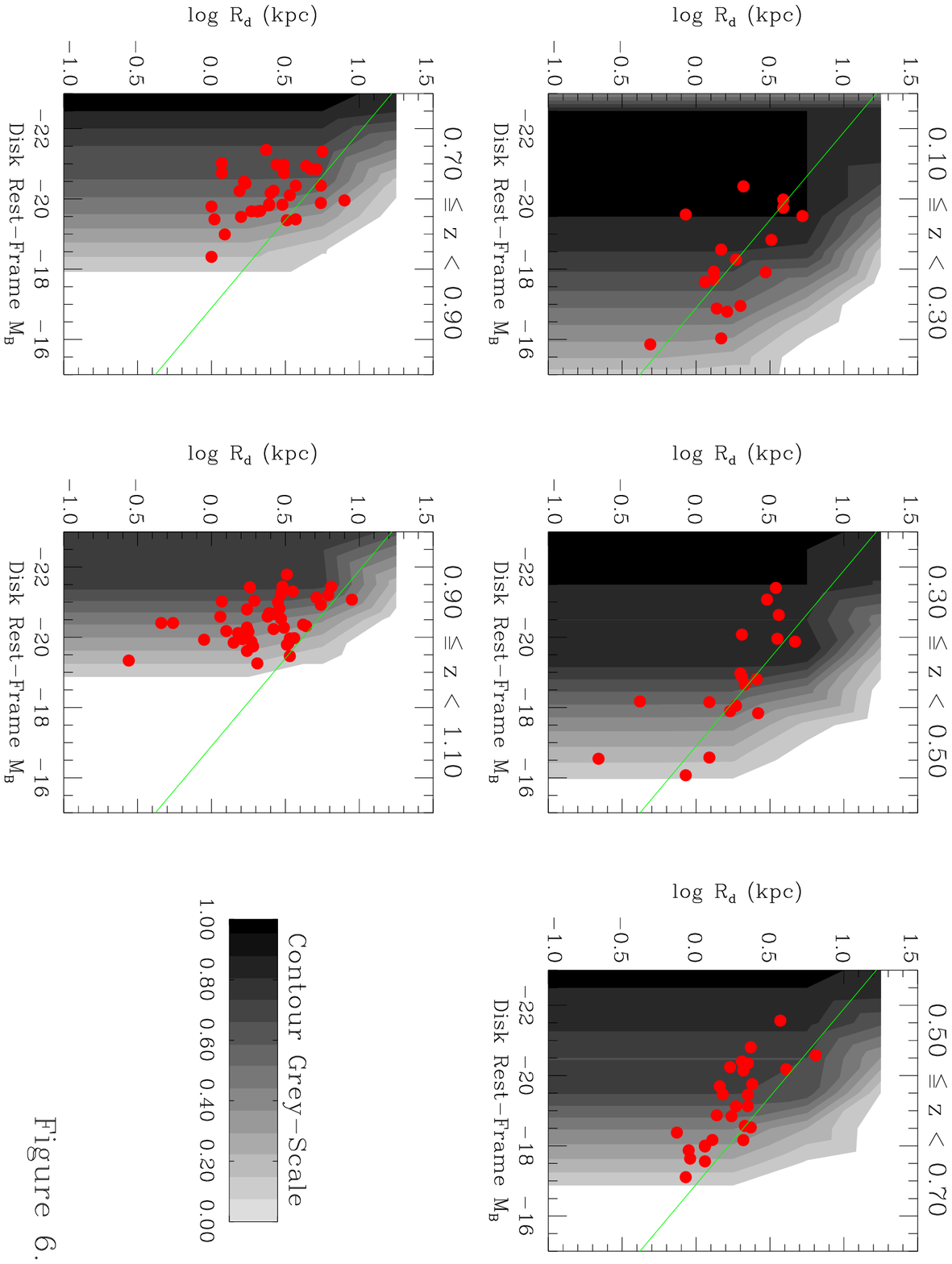]{Rest-frame selection function 
$S_{US}$($M_{B_{0}}$, $R_{d}$) ({\it shaded contours}) as a function of 
redshift.  The highest contour was normalized to one in each redshift 
bin.  Red symbols are galaxies from Figure~\ref{sizemag-disk} with 
Keck redshifts and $B/T < 0.5$, and the solid green line is the 
canonical Freeman relation. 
\label{rest-selecf}}

\figcaption[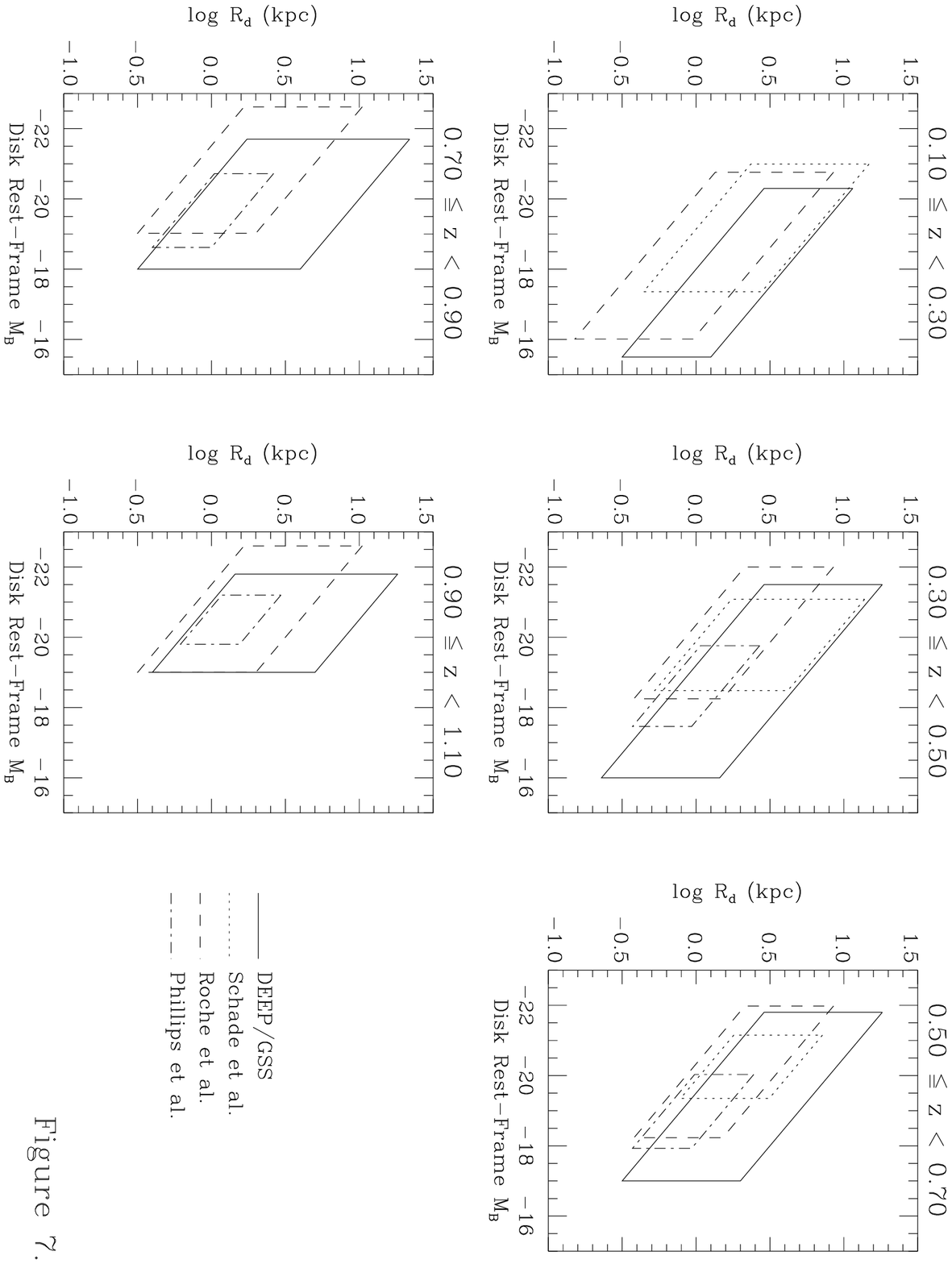]{Loci of different magnitude-size samples: 
DEEP/GSS ({\it solid outline}), Schade et al.  (1996b) ({\it dotted 
outline}), Roche et al.  (1998) ({\it dashed outline}), and Phillips 
et al.  (1997) ({\it dot-dashed outline}).  The DEEP/GSS 
outlines are the same as those shown in Figure~\ref{sizemag-disk}. 
Note that the DEEP/GSS sample penetrates to lower surface 
brightnesses at high redshift than the other samples.
\label{compare-loci}}

\figcaption[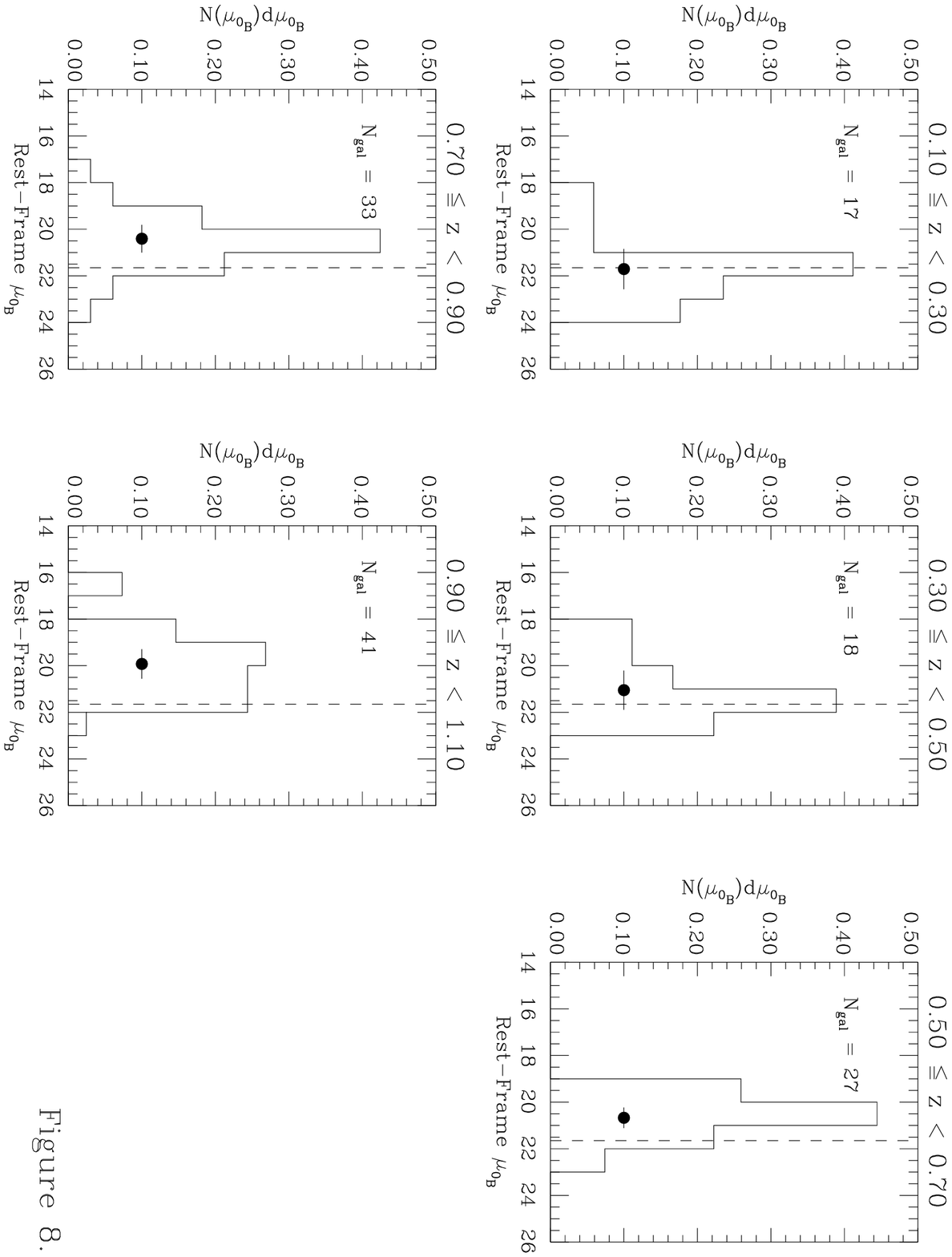]{Raw disk rest-frame B-band central surface 
brightness distributions as a function of redshift uncorrected by the 
survey selection function $S_{US}$.  Surface brightnesses were 
computed using Equation~\ref{mu0}.  The area under each distribution 
was normalized to one, and the vertical dashed lines mark the value of 
the canonical Freeman disk central surface brightness ($\mu_{0_{B}} = 
21.65$ mag arcsec$^{-2}$).  The large filled circles give the location 
of the mean surface brightness in each redshift bin.  The error bar on 
the location of the mean is the 3-sigma standard error.  The mean 
surface brightness increases systematically from $\mu_{0_{B}} = 21.7$ 
mag arcsec$^{-2}$ at $z = 0.2$ to $\mu_{0_{B}} = 20.4$ mag 
arcsec$^{-2}$ at $z = 0.80$, similar to that seen by Schade et al.  
and Roche et al.
\label{rest-sb-histo-nosel}}

\figcaption[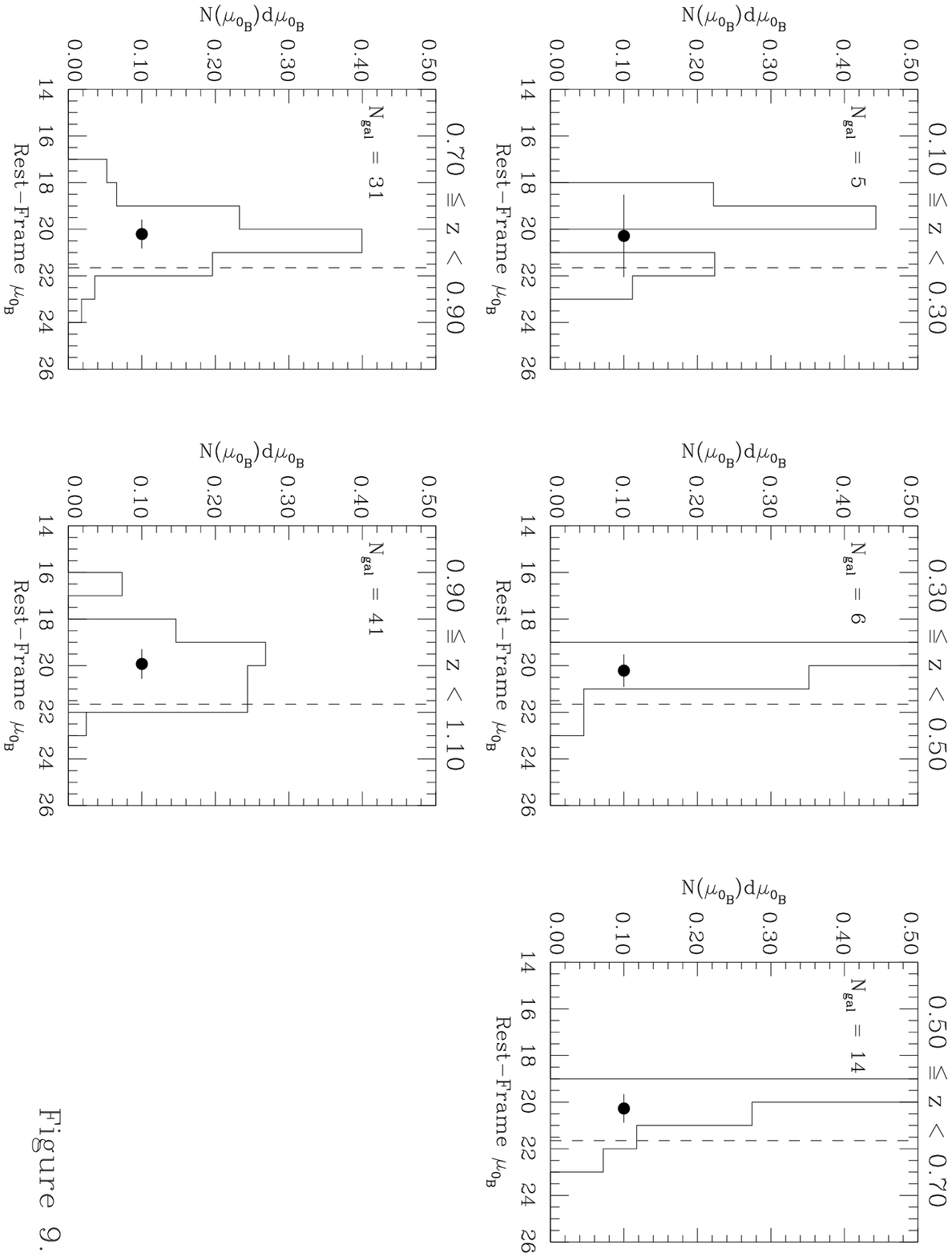]{Adjusted disk rest-frame B-band central surface 
brightness distributions as a function of redshift scaled by the 
survey selection function $S_{US}$ following Equation~\ref{mu0dist}. 
This scaling converts each distribution to what it would look like if 
observed with the same selection function as the highest redshift bin. 
Each galaxy thus contributes the fraction ${{S_{US}(R_{d_{ijk}}, 
M_{ijk}, 0.90 \leq z)}/{S_{US}(R_{d_{ijk}}, M_{ijk}, z_{i})}}$ to 
these adjusted distributions.  $N_{gal}$ is the number of galaxies in 
each bin for which $S_{US}(R_{d_{ijk}}, M_{ijk}, 0.90 \leq z) \neq 
0$.  The vertical dashed lines mark the value of the canonical Freeman 
disk central surface brightness ($\mu_{0_{B}} = 21.65$ mag 
arcsec$^{-2}$).  The large filled circles give the location of the 
mean surface brightness in each redshift bin.  The error bar is the 
3-sigma standard error on the mean.  There is no systematic increase 
in the mean surface brightness from $z = 0.2$ to $z = 0.8$.  This 
shows how important selection effects can be in interpreting the 
magnitude-size relation at high redshift.  A significant fraction 
($\sim$ 25$\% $) of galaxies in the highest redshift bin remained 
close to the Freeman canonical surface brightness of $\mu_{0_{B}} = 
21.65$ mag arcsec$^{-2}$.
\label{rest-sb-histo-sel}}

\figcaption[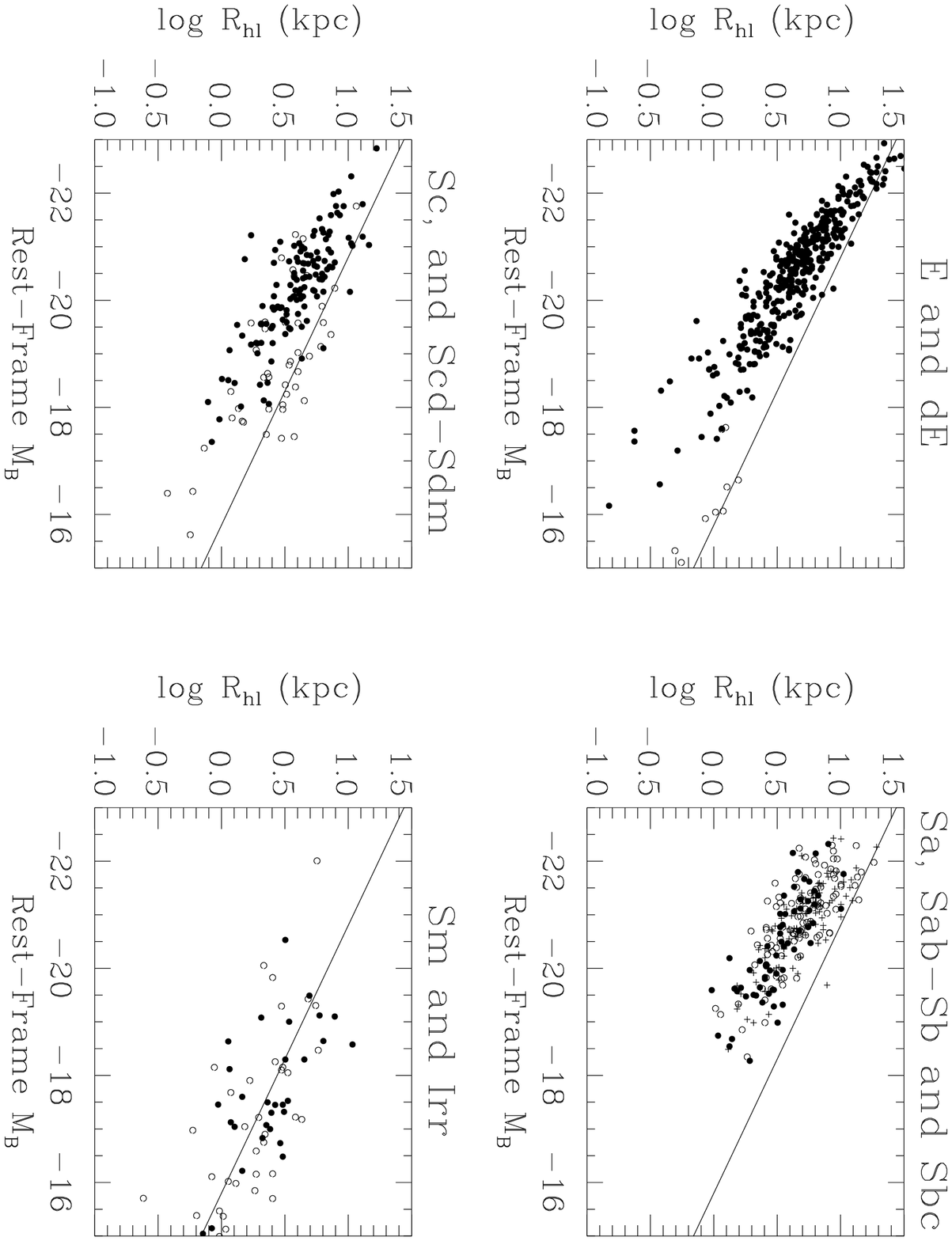]{Half-light radius and total rest-frame B 
magnitude data for 957 galaxies taken from Burstein et al.  (1997) and 
Bender et al.  (1992).  Upper left-hand panel: ellipticals E ({\it 
filled circles}) and dwarf ellipticals dE ({\it open circles}).  Upper 
right-hand panel: Sa ({\it filled circles}), Sab-Sb ({\it open 
circles}) and Sbc ({\it pluses}).  Lower left-hand panel: Sc ({\it 
filled circles}) and Scd-Sdm ({\it open circles}).  Lower right-hand 
panel: Sm ({\it filled circles}) and Irr ({\it open circles}).  The 
solid line is the half light radius-magnitude relation for a canonical 
Freeman disk with a central surface brightness of 21.65 mag 
arcsec$^{-2}$.  The Freeman relation is provided for the E galaxies 
only as a reference since, strictly speaking, it applies only to disk 
galaxies.
\label{local-kappa}}

\figcaption[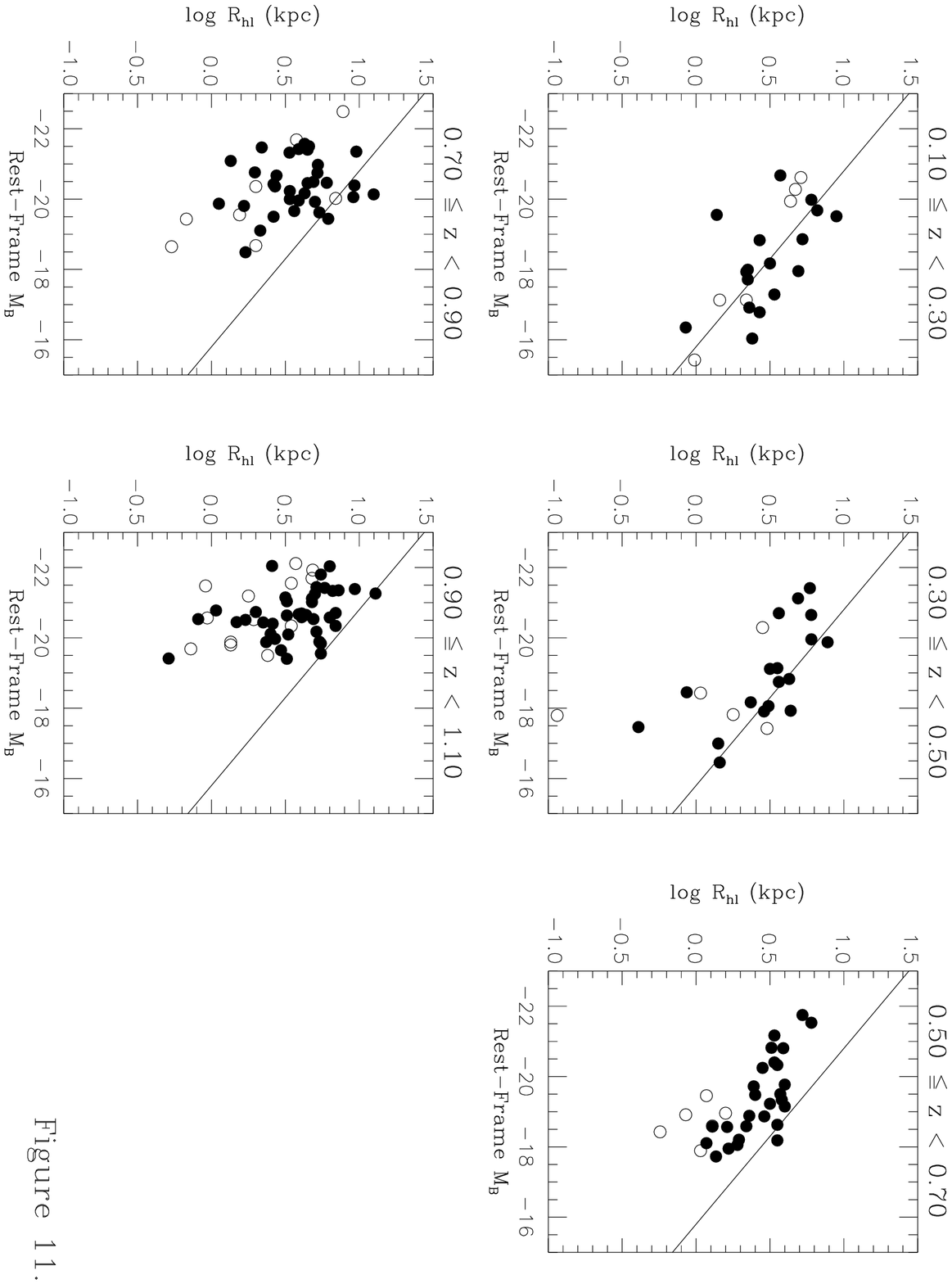]{Half-light radius in kiloparsecs 
versus total rest-frame B-band absolute magnitude as a function of 
redshift for all galaxies with Keck redshifts.  {\it Filled circles:} 
$0.0 \leq B/T < 0.5$, {\it Open circles:} $0.5 \leq B/T \le 1.0$. The 
solid line is the half-light radius-magnitude relation for a canonical 
Freeman disk with a central surface brightness of 21.65 mag 
arcsec$^{-2}$.
\label{sizemag-all}}

\figcaption[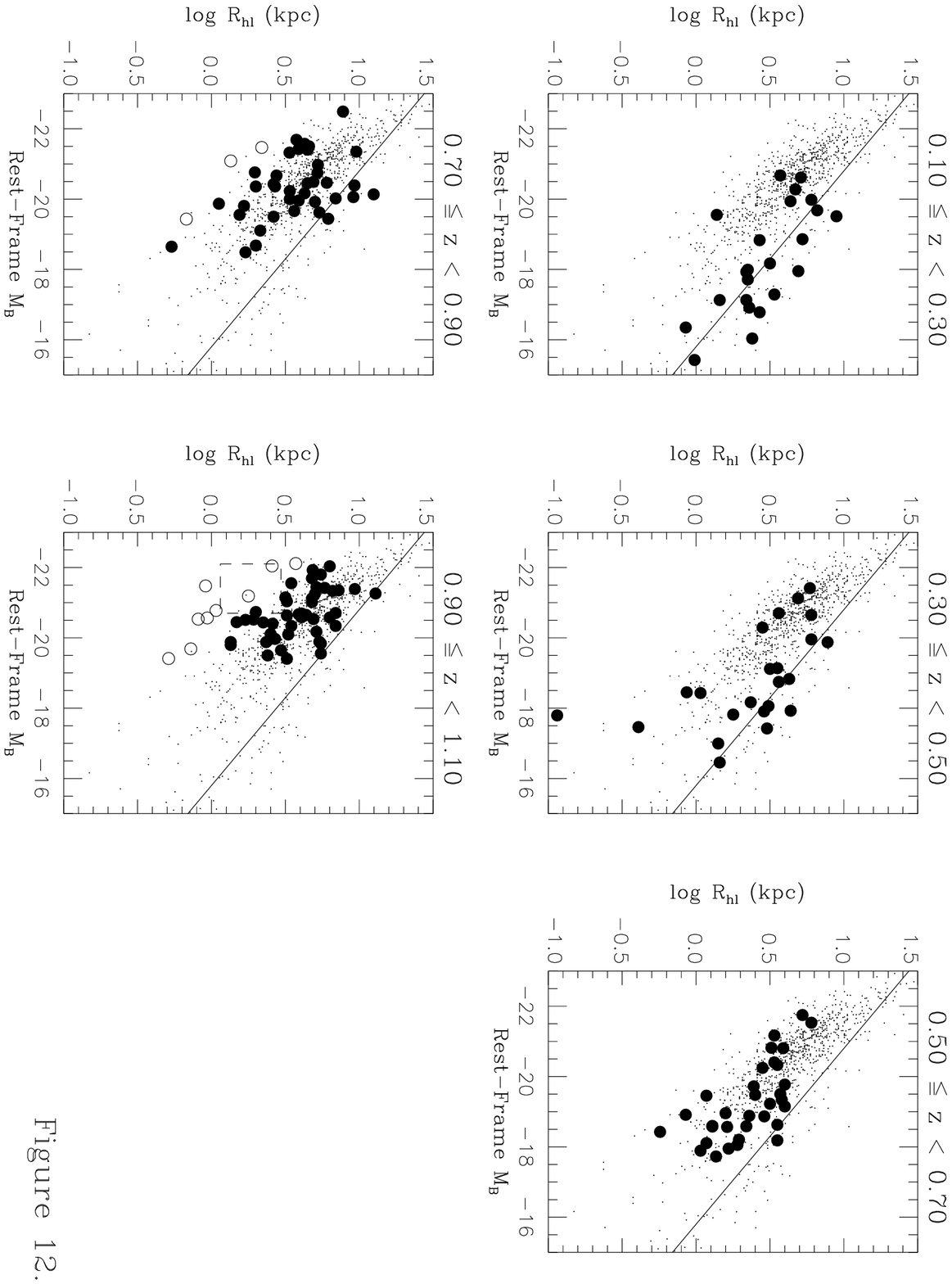]{Direct half-light radius-magnitude comparison 
between local BBFN galaxies (dots) and galaxies in the DEEP/GSS 
redshift sample (filled circles).  The HSB galaxies are shown as open 
circles.  The dashed box outlines the magnitude-size locus of the $z 
\sim 3$ galaxies from the sample of Lowenthal et al.  (1997), and the 
solid line is the Freeman relation.  Comparison with 
Figure~\ref{local-kappa} shows that the DEEP/GSS sample probably 
includes different morphological mixtures in different redshift bins, 
ranging from Sm-Irr galaxies at low redshift to early-type and HSB 
galaxies at high redshift.
\label{sizemag-local-distant}}

\clearpage
\begin{deluxetable}{ccccccc}
\tablecaption{Rest-Frame Surface Brightness Distribution Parameters}
\tablehead{
\colhead{} & \multicolumn{3}{c}{No Selection Function Adjustment} & 
\multicolumn{3}{c}{With Selection Function Adjustment}\nl 
\colhead{$z$} & \colhead{$N_{gal}$} & \colhead{$\overline{\mu}_{0_{B}}$} & 
\colhead{$\sigma(\mu_{0_{B}})$} & \colhead{$N_{gal}$} & 
\colhead{$\overline{\mu}_{0_{B}}$} & \colhead{$\sigma(\mu_{0_{B}})$}
}
\startdata
$0.10 \leq z < 0.30$  & 17 & 21.7 $\pm$ 0.3 & 1.2 & 5 & 20.3 $\pm$ 0.6 & 1.3 \nl
$0.30 \leq z < 0.50$  & 18 & 21.0 $\pm$ 0.3 & 1.2 & 6 & 20.2 $\pm$ 0.2 & 0.6 \nl
$0.50 \leq z < 0.70$  & 27 & 20.7 $\pm$ 0.2 & 0.8 & 14 & 20.3 $\pm$ 0.2 & 0.8 \nl
$0.70 \leq z < 0.90$  & 33 & 20.4 $\pm$ 0.2 & 1.1 & 31 & 20.2 $\pm$ 0.2 & 1.2 \nl
$0.90 \leq z < 1.10$  & 41 & 19.9 $\pm$ 0.2 & 1.4 & 41 & 19.9 $\pm$ 0.2 & 1.4 \nl

\enddata
\label{sb_histo_param}
\end{deluxetable}

\clearpage
\appendix
\section{APPENDIX A: Rules for the Spectroscopic Selection Function $S_{PS}$} \label{appA}

Ideally, $S_{PS}(I_{814},r_{d})$ should just be the ratio of the 
spectroscopic and photometric samples binned in both magnitude {\it 
and} size.  Unfortunately, this is not always the case for two 
reasons.  First, even with large bin sizes ($\Delta mag = 1.0$ and 
$\Delta$ log $r_{d}$ = 0.5), small-number statistics at the bright end 
cause large fluctuations in $S_{PS}(I_{814},r_{d})$ where it should 
have been flat and close to its maximum value.  To alleviate this 
small-number problem, the spectroscopic and photometric samples were 
binned only in magnitude ($\Delta mag = 1.0$).  The resulting 
$S_{PS}(I_{814},r_{d})$ selection function is thus insensitive to any 
size-dependent effect in the spectroscopic selection.  This would be a 
problem if the size distributions of galaxies in the spectroscopic 
sample were markedly different from those of the general population as 
a function of magnitude.  No difference was found in four magnitude 
ranges: $20.5 \leq I_{814} < 21.5$, $21.5 \leq I_{814} < 22.5$, $22.5 
\leq I_{814} < 23.5$, and $23.5 \leq I_{814} < 24.5$.  Second, the 
ratio $\Psi_{S}/\Psi_{P}$ of the spectroscopic and photometric samples 
must be calculated in each bin in the $I_{814}-$log $r_d$ plane 
according to the rules shown in Table~\ref{sgsrules}.  
Table~\ref{sgsrules} describes all the different ways in which the 
selection functions operate on $\Psi_{U}$ and $\Psi_{P}$ to produce 
$\Psi_{S}$.  ``Y'' and ``N'' are used to indicate whether galaxies in 
the magnitude-size bin are present or not in the $\Psi$'s.  For 
simplicity, zeroes and ones are used for the selection functions.  In 
reality, they can obviously take on any value between 0 and 1 since 
they give the probabilities of detecting objects photometrically and 
spectroscopically.

Cases (3) and (7) are the only two cases where $S_{PS}$ is greater 
than $S_{UP}$.  These cases represent serendipitous observations of 
very strong emission-line objects which are not detected on 
broad-band images.  Since these objects were not included in the 
spectroscopic sample used for this paper, Cases (3) and (7) do not 
apply here.  $S_{PS}$ must be carefully interpreted when both 
$\Psi_{P}$ and $\Psi_{S}$ are zero.  Take Case (5) for example.  Objects 
in this case are absent from both $\Psi_{P}$ and $\Psi_{S}$ because 
they do not exist in Universe and not because they could not have been 
detected in principle.  It would therefore be wrong to set $S_{PS}$ to 
zero if these selection functions were to be applied to predictions 
from theoretical models.  The selection functions should not 
erase potential differences between theoretical model predictions and the 
true galaxy population in the Universe.  The best estimate for 
$S_{PS}$ in this case is therefore an upper limit given by the 
value of $S_{UP}$ in the same magnitude-size bin. Therefore, $S_{PS}$ 
was set to $S_{UP}$ whenever both $\Psi_{P}$ and $\Psi_{S}$ were zero. 
Otherwise, $S_{PS}$ was simply set equal to $\Psi_{S}$/$\Psi_{P}$.

The selection rules discussed so far in this Appendix apply to the 
very general case of a survey in which redshifts may not have been 
obtained for all the objects observed and in which redshifts may have 
been obtained serendipitously for objects absent from its photometric 
catalog.  The DEEP/GSS sample ($I_{814} \leq 23.5$) used in this paper 
was much simpler to analyze than such a general survey since Keck 
redshifts were obtained for 100$\%$ of the objects in the sample, and 
all objects were present in the DEEP/GSS photometric catalog.  In this case, 
the spectroscopic selection function $S_{PS}$ in a given 
($I_{814},r_{d}$) bin is simply the ratio of the number of galaxies 
observed spectroscopically and the number of galaxies in the 
photometric sample.

\clearpage
\begin{deluxetable}{cccccc}
\tablewidth{0pt}
\tablecaption{Spectroscopic Selection Function $S_{PS}$ Rules}
\tablehead{
\colhead{Case} & \colhead{Universe} & \colhead{$S_{UP}$} & 
\colhead{Photometric Catalog} & 
\colhead{$S_{PS}$} & \colhead{Spectroscopic Catalog} \nl
\colhead{} & \colhead{$\Psi_{U}$} & \colhead{} & 
\colhead{$\Psi_{P}$} & 
\colhead{} & \colhead{$\Psi_{S}$}
}
\startdata
1 & Y & 1 & Y & 1 & Y \nl
2 & Y & 1 & Y & 0 & N \nl
3 & Y & 0 & N & 1 & Y \nl
4 & Y & 0 & N & 0 & N \nl
5 & N & 1 & N & 1 & N \nl
6 & N & 1 & N & 0 & N \nl
7 & N & 0 & N & 1 & N \nl
8 & N & 0 & N & 0 & N \nl
\enddata
\label{sgsrules}
\end{deluxetable}

\end{document}